\documentstyle[12pt]{article}
\newcommand{\Closure}{\mbox{${\large C}$}}
\newcommand{\ClosCon}{\mbox{${\large D}$}}
\newcommand{\Conicalone}{\mbox{${\large D}_1$}}
\newcommand{\Conicaltwo}{\mbox{${\large D}_2$}}
\newcommand{\Conicalthree}{\mbox{${\large D}_3$}}
\newcommand{\Yield}{\mbox{$\Upsilon$}}
\newcommand{\sign}{\hbox{\rm sign}}
\newcommand{\CO}{\mbox{$c_{0}$}}

\newcommand{\szz}{\mbox{$\sigma_{zz}$}}
\newcommand{\gzz}{\mbox{$s_{zz}$}}
\newcommand{\srr}{\mbox{$\sigma_{rr}$}}
\newcommand{\sxx}{\mbox{$\sigma_{xx}$}}

\newcommand{\sxz}{\mbox{$\sigma_{xz}$}}
\newcommand{\grr}{\mbox{$s_{rr}$}}
\newcommand{\szr}{\mbox{$\sigma_{zr}$}}
\newcommand{\srz}{\mbox{$\sigma_{rz}$}}
\newcommand{\gzr}{\mbox{$s_{zr}$}}
\newcommand{\grz}{\mbox{$s_{rz}$}}
\newcommand{\scc}{\mbox{$\sigma_{\chi \chi}$}}
\newcommand{\gcc}{\mbox{$s_{\chi \chi}$}}
\newcommand{\src}{\mbox{$\sigma_{r \chi}$}}
\newcommand{\scz}{\mbox{$\sigma_{\chi z}$}}
\newcommand{\szc}{\mbox{$\sigma_{z \chi}$}}
\newcommand{\scr}{\mbox{$\sigma_{\chi r}$}}
\newcommand{\snn}{\mbox{$\sigma_{nn}$}}
\newcommand{\sss}{\mbox{$\sigma_{mm}$}}
\newcommand{\sns}{\mbox{$\sigma_{nm}$}}

\newcommand{\aaa}{\mbox{$\hat{a}$}}
\newcommand{\bbb}{\mbox{$\hat{b}$}}
\newcommand{\ccc}{\mbox{$\hat{c}$}}
\newcommand{\ddd}{\mbox{$\hat{d}$}}
\newcommand{\eee}{\mbox{$\hat{e}$}}
\newcommand{\fff}{\mbox{$\hat{f}$}}

\begin{document}

\title{Stress Propagation and Arching in Static Sandpiles}

\author{J.~P.~Wittmer and M.~E.~Cates\\
Department of Physics and Astronomy\\   University of Edinburgh, JCMB King's
Buildings\\ Mayfield Road, Edinburgh EH9 3JZ, UK\\
\\ P. Claudin$^*$\\Cavendish Laboratory, Madingley Road\\  Cambridge
CB3 OHE, UK\\ $^*$ Present Address:
\\ Service de Physique de l'Etat Condens\'e, CEA\\
Ormes des Merisiers, 91191 Gif-sur-Yvette, Cedex France
}

\date{1 May 1996}
\date{}
\setcounter{page}{0}
\maketitle
\thispagestyle{empty}
\begin{center}
PACS numbers 46.10.+z, 46.30.-i, 81.35.+k, 83.70 Fn
\end{center}
\newpage

\begin{abstract}
We present a new approach to the modelling of stress
propagation in static granular media, focussing on the conical sandpile
constructed from a point source.
We view the medium as consisting of cohesionless
hard particles held up by static frictional forces; these are subject to
microscopic indeterminacy which corresponds macroscopically to the fact that the
equations of stress continuity are incomplete -- no strain variable can be
defined.  We propose that in general the continuity equations
should be closed by means of a constitutive relation (or relations) between
different components of the (mesoscopically averaged) stress tensor.
The primary constitutive relation relates radial and vertical shear and
normal stresses (in two dimensions, this is all one needs).  We argue that the
constitutive relation(s) should be local, and should {\it encode the
construction history of the pile}: this history determines the
organization of the grains at a mesoscopic scale, and thereby the
local relationship between stresses.
To the accuracy of published experiments,
the pattern of stresses beneath a pile shows a scaling between
piles of different
heights (RSF scaling) which severely limits the form the constitutive relation
can take; various asymptotic features of the stress patterns can be
predicted on the basis of this scaling alone. To proceed further, one
requires an
explicit choice of constitutive relation; we review some from the literature
and present two new proposals. The first, the FPA (fixed principal axes) model,
assumes that the eigendirections (but not the eigenvalues) of the stress tensor
are determined forever when a material element is first buried.
(This assumes, among other things, that subsequent loadings are not so large as
to produce slip deep inside the pile.) A macroscopic consequence of this
mesoscopic assumption is that the principal axes have fixed orientation {\it in
space}: the major axis everywhere bisects the vertical and the free surface. As
a result of this, stresses propagate along a nested set of archlike structures
within the pile, resulting in a {\it minimum} of the vertical normal stress
between the apex of the pile, as seen experimentally (``the dip"). This
experiment has not been explained within previous continuum approaches; the
appearance of arches within our model corroborates earlier physical arguments
(of S. F. Edwards and others) as to the origin of the dip, and places them on a
more secure mathematical footing. The second model is that of ``oriented stress
linearity" (OSL) which contains an adjustable parameter (one value of
which corresponds to FPA). For the general OSL case, the simple
interpretation in
terms of nested arches does not apply, though a dip is again found over a
finite parameter range. In three dimensions, the choice for the primary
constitutive relation must be supplemented by a secondary one; we have tried
several, and find that the results for the stresses in a three dimensional
(conical) pile do not depend much on which secondary closure is chosen. Three
dimensional results for the FPA model are in good
semiquantitative agreement with published experimental data on conical piles
(including the dip); the data does not exclude, but nor does it support,
OSL parameters somewhat different from FPA. The modelling strategy we adopt,
based on local, history-dependent constitutive relations among stresses,
leads to nontrivial predictions for piles which are prepared with a different
construction history from the normal one. We consider several such histories in
which a pile is  prepared and parts of it then removed and/or tilted.
Experiments
along these lines could provide a searching test of the theory.

\end{abstract}


\newpage

\section{Introduction}

\label{Introduction}

A sandpile is normally constructed by pouring sand from
a stationary point source, as shown in Fig.1(a).
Each element of sand arrives at the apex of the pile, rolls down the slopes,
comes to rest, and is finally buried. The final (static) sandpile then consists
of a symmetrical cone whose surface is at the angle of repose of
the material. Some of the simplest questions one can ask about
this system concern the distribution of stresses in the pile.
Specifically, it is possible experimentally to measure the downward force
on the supporting surface at different positions under the pile
\cite{jokati,smid}. (Throughout the paper we assume this to be a high friction
surface so that slip does not occur at it.) Intuitively one can guess that the
maximum force would be recorded directly beneath the apex of the pile; but in
fact, the experiments show a pronounced {\em dip} in the force beneath the apex.
This counterintuitive result has stimulated
various theoretical \cite{Hong,Huntley,EO,EM,BCC}
and computational \cite{bagster,liffmann} studies; but so far, there has
been no clear
consensus on the origin of the dip.

In the present work we pursue a continuum mechanics approach based on the
equations of stress continuity in a cohesionless granular
medium \cite{nedderman,BCC,sokolovskii,drescher}. This approach immediately
encounters the
problem of {\em indeterminacy}: even in the simplest case of a two-dimensional
pile (which we consider in detail), the continuity of stress does not lead to a
closed set of equations. In elastic materials, this deficiency would be
rectified by invoking the usual {\em constitutive relation} between stress and
strain (Hooke's law). For granular media, however, there is no clear definition
of ``strain". Rather, it is widely assumed that the physics of granular media
can be understood purely in terms of rigid particles packed together in
frictional contact (so that no strain variables can be defined). The
indeterminacy of the stress equations then has a clear origin: for two rigid
particles in frictional contact with a specified normal force, the coefficient
of static friction defines only the {\em maximum} shear force that may be
present. Our continuum mechanics approach, like some previous ones
\cite{BCC} assumes that, despite this local indeterminacy, there
emerges on length scales much larger than the grain size some definite relation
between the {\em average} frictional and normal forces. Thus we assume the
existence of one or more constitutive relations, not between stress and strain,
but among the various components of the stress tensor itself. In two dimensions,
one such relation is enough to close the equations; in three dimensions (subject
to certain symmetry assumptions) two are needed.

A basic tenet of our
approach is that the constitutive relations are {\it local}: we assume that
that these relations between stresses do not depend on distant
perturbations, although the
stresses themselves certainly do.
Clearly, the constitutive relation (or relations -- we suppress the plural in
what follows), between stresses in some material element, must reflect the
packing arrangement of grains in that element.
This raises the possibility that
the constitutive behaviour could vary from place to place in the pile. More
generally, we believe that in principle the constitutive relation of a given
material element should encode its entire {\em construction history}.
We shall
assume, however, that  the important part of this history comes to an end at the
moment where the element is buried: at later times, although the stresses
passing
through the element may vary, the  constitutive relation
between them cannot. We call this the assumption of {\em
perfect memory}. We show below (Sec. 2.3) that the perfect memory and
locality assumptions, when
combined with a simple and experimentally motivated scaling assumption (called
RSF scaling),  drastically limit the form the constitutive equation can take.

A consequence of perfect memory is that the
stresses in an element, once buried, respond reversibly (though
not necessarily linearly) to any subsequent additional loading. Such loadings
can be brought about either by adding more material to the pile, or by putting a
small weight on its surface, for example.  Obviously our perfect memory
assumption, and indeed that of locality, may fail if the load added is so
large as to lead to rearrangement of grains within the element (that is, slip).
>From our viewpoint, however, if slip does occur, this represents a change in
construction history which must explicitly be taken into account. It turns out
that for most of the models and geometries considered in this paper,
perturbative loadings of the pile do not, in fact, cause slip except
at the surface of the pile.

Our assumption of a local, history-dependent constitutive relation among
stresses is not widely
accepted as a modelling strategy for sandpiles. (Indeed, we have not seen
any really clear exposition of this strategy in the previous literature.)
Many would argue the necessity of explicitly invoking the deformability of
particles (allowing a strain variable to enter);  others would argue that
infinitesimal distant loads should cause rearrangement of a network of contacts
among hard particles (leading to intrinsically nonlocal stress propagation).
This paper aims to explore in detail
the kinds of prediction that can be made within our overall modelling strategy,
and to introduce some physically plausible candidates for constitutive
relations.

\subsection{Previous continuum approaches}

Much existing work on the static continuum mechanics of cohensionless granular
materials has invoked, as an implicit constitutive relation, the assumption of
{\em Incipient Failure Everywhere} (IFE). That this is indeed an
{assumption},  is not always made clear in
the engineering literature \cite{nedderman,sokolovskii,drescher,Wood}. The
IFE model
supposes that the material is everywhere just on  the point of slip failure: all
frictional forces are ``fully mobilized''. Thus an appropriately chosen (local) {\em
yield criterion} \cite{nedderman,drescher} of the
material provides the missing constitutive equation.

The physics of this assumption is dubious. When a pile is made
from a point source (Fig.1(a)) there is a continuous series of
landslides at the surface; we therefore accept that an incipient failure
condition is maintained {\em at the surface}. Even in saying this, we ignore the
distinction between the angle of repose (that of the free surface just
after a landslide) and the maximum angle of stability (that just before). These
differ by the Bagnold hysteresis angle, which is small compared to the repose
angle itself \cite{Bagnold,BCPE}; we neglect this hysteresis effect from now
on. However, the validity of the incipient failure condition at the surface,
which contains material elements just at the point of burial, does not mean that
the same condition still  holds for an element long afterwards. Such an element
lies deep beneath the surface, and has since burial been loaded by adding more
material to the pile above. In fact, the  IFE assumption can clearly be ruled
out on experimental grounds: as we show below, it fails to
account for the dip.


The IFE closure
implies that two of the {\em principal stresses} are proportional
\cite{nedderman}. In contrast, proportionality of the {\em horizontal and
vertical normal stresses} was proposed as a constitutive relation
recently by Bouchaud, Cates and Claudin \cite{BCC}  (we refer to this closure as
the BCC model). In two dimensions it was found that the stress continuity
equation then has a convenient analytic property: it becomes a wave equation.
However \cite{BCC}, the BCC model predicts a stress
plateau, rather than a dip, at the centre of a sandpile. Attempts to explain the
dip by introducing various nonlinear  terms in the constitutive relation proved
unconvincing, at least for small nonlinearities, whose perturbative inclusion
showed a {\em hump} instead of a dip. The BCC paper (Sec.3.2 of Ref. \cite{BCC})
in fact included a brief discussion of certain strongly nonlinear models
\cite{language}, which
were also argued to give a hump. This conclusion
turns out to be incorrect, for subtle reasons that we discuss
below (Sec. 2.9).  In one sense, the ``oriented stress linearity" (OSL)
model, which we
study in detail in this paper, can be viewed as an extreme limit of
this type of model. The OSL model,
like BCC, has a {\em linear relation between normal stresses} but now in a
coordinate system that is tilted (through a  constant angle) with respect
to the vertical.


An apparently completely different approach to describing
the stress distribution in sandpiles was proposed by
Edwards and Oakeshott \cite{EO}. These authors considered a
pile consisting of a stack of nested {\em arches}, Fig.1(b).
(A recent modification considers a model of
platelike granules and allows curvature of the arches
\cite{EM}).  Each arch supports only its own
weight, and consequently the vertical stress decreases for the
smaller arches near the centre of the pile.
This approach provides a very appealing physical picture
of why there is a dip. In the arching
mechanism, the load in an
element is transmitted unevenly to those below. The central part of the pile is
thereby ``screened" from additional loadings which are supported instead
by the outer regions.

However, there are some obvious drawbacks with this approach. Firstly, the dip
is greatly {\em overpredicted}: by construction there is no downward force
whatever at the centre of the pile, while experiments show a finite value.
Secondly, unless the arches are parallel to the free surface, the outermost
``arches" are incomplete. It is mechanically impossible for one of these to
transmit its weight purely along its own length: there is an unbalanced couple
about the base of such an arch which would cause it to fall over. If, instead,
the arches are parallel to the free surface then the model is stable, but it
predicts an abrupt discontinuity in the downward force at the edge of the pile,
which is not observed.

These difficulties
arise, at least in part, from an inconsistent attempt to treat the vertical
normal stress (``weight") independently of the other stress components.
This is rectified in the OSL models introduced below.  Among these is
a special case, the ``fixed principal axes" (FPA) model,
which is very close in its physical content to the picture of nested arches
originally suggested by Edwards and Oakeshott. As its name suggests, in the
FPA model,
the principal axes of the stress tensor have a constant angle of inclination to
the vertical. These axes turn out to coincide with
the stress-propagation characteristics, which
resemble a set of Edwards arches (Sec. 2.7). In fact, we believe that our FPA
model gives, for the first time, a fully consistent continuum mechanics
implementation of Edwards' arching picture. As shown in Sec. 3, this model
gives good agreement with experimental data in three-dimensional
sandpiles.

The discussion above (like others in the recent physics literature on
sandpiles) attributes the idea of arching to Edwards and Oakeshott \cite{EO}.
However, the same basic picture has a longer pedigree in the rock mechanics
literature, and can be traced at least as far back as the pioneering work of
Trollope in the 1950's
\cite{Trollope50s}. (The latter is very clearly reviewed in \cite{Trollope}.)
Something very like the Edwards-Oakeshott model is called by Trollope the ``full
arching limit" and something very like the BCC
model turns out to be the ``no arching limit" of Trollope's theory (although
his predictions based on the latter do not take proper account of the Coulomb
yield criterion, as was done by BCC). Trollope also developed a ``systematic
arching theory" to provide an interpolation between these two limits.
This model, though it does not provide a systematic theory of
arching, is quite interesting, and we discuss it in more detail in Section 2.10.
It is based on quite different physical principles from our own work, partly
because Trollope attributed the arching phenomenon to {\it small displacements
of the supporting surface} under a wedge. This idea, based on Trollope's own
experiments (on wedges whose construction history we have been unable to find
out) is clearly at odds with our own explanation, and appears to be contradicted
by the more recent experimental data on sandpiles constructed from a point
source \cite{jokati,smid,deflection}.

\subsection{Related modelling work}

Other approaches to the problem of stress propagation in static sandpiles
include
particle packing models \cite{Hong,Huntley}, where one considers a regular
packing of (usually spherical) grains, with simple transmission laws for
the downward force between one layer of particles and those below.
These models show a flat stress plateau in two dimensions,  a feature shared
with  BCC \cite{BCC} which can be viewed as a (slightly generalized) continuum
limit of such models.
An important and related class of discrete models address the propagation of
{\em noise effects} in sandpiles; in these the transmission of forces between
particles is stochastic \cite{Liu,degennes}. The relation between these models
and our own (noise-free) continuum approach will be explored in detail elsewhere
\cite{ournoise}.

More elaborate discrete models are increasingly being studied. That examined
numerically by Bagster and Kirk \cite{bagster} invokes nontrivial
force propagation rules locally, and for some parameter ranges shows
a dip in the stress. However, it is not clear whether the physics included
in this model is that of real sandpiles; and so far, the
relation to any continuum description is uncertain.
A widespread numerical approach is that of discrete element modelling
\cite{discrete}; however,
as discussed by Buchholtz and Poeschel \cite{volkard1}, many such methods cannot
so far even reproduce the fact that the repose angle of a pile is independent
of its size. Various improved algorithms have been suggested \cite{goddard};
we do not know whether the same is true for these.
In a future paper \cite{volkard2} we will present results from
a simulation approach involving nominally frictionless, but nonspherical,
slightly deformable particles, following Ref.\cite{volkard1}. These offer
some promise
of confirming, or at least testing,
the ideas put forward in the present work.

The model of Liffman {\em
et al.} \cite{liffmann} invokes a more specific mechanism, based
on size-segregation, to explain the dip in the stress. When confronted with the
experimental data
\cite{jokati,smid} one sees a serious drawback of this explanation: the
data show a scaling behaviour which
the model cannot support. The observed scaling (called RSF scaling, see
Sec. 2.3 below) indicates that there is {\em no characteristic length-scale}
intrinsic to the granular medium of which a pile is made. In general,
segregation effects introduce such a scale by setting up gradients in the
material properties of the pile \cite{footdistrib}, and hence violate the
observed scaling. It is notable that a finite deformability of the particles
would also introduce a characteristic length, and is therefore also ruled
out by RSF scaling (we discuss this further in Sec. 2.3).

The IFE model, defined above, represents one limiting case of a more
general group of elasto-plastic continuum theories; some of these are
highly developed and widely used
within a finite element framework (though usually in the context of hoppers
rather than sandpiles) \cite{karlsruhe}. The physical basis of these
models for dry cohesionless granular media is not always clear
(many are based on models developed earlier for wet soil \cite{Wood}). In any
case, to whatever extent elasticity is invoked, such models are again in
violation of RSF scaling.

The idea that the properties of a granular medium depend on its construction
history is central to our work. This concept is not new, and plays a strong
role in the recent experimental
literature on granular media in hoppers (for example the exit flow from
a hopper depends on how it was filled) \cite{rotter1}. Indeed, this
is part of the reason why standardized shear and triaxial tests are used to
measure the internal friction coefficient of a granular medium; the repose
angle, which in the simplest theories is completely equivalent
\cite{nedderman}, in reality depends appreciably on construction history,
as do other
mechanical properties \cite{rotter1} (this is discussed further in Section 4).
Moreover, it has long been argued that the manner in which the construction
history enters is via the local packing geometry
of the grains. This forms part of the idea of ``granular fabric", in which one
constructs a local tensor that parameterizes the
distribution of particle-particle contacts \cite{fabric}. The concept
of the fabric tensor has usually been developed in an elasticity context,
rather than one in which constitutive relations directly among stresses
are assumed. Nonetheless,
the orientational memory effects embodied in the new models described below
can certainly be viewed in terms of a local tensorial property of the
medium (see Sec. 2.7). However, no specific interpretation of this
quantity (in terms of the contact distribution) appears to be required.

\subsection{The present work}

In Section \ref{sec2D} we give a coarse-grained continuum description
of the two-dimensional symmetrical sandpile.
Instead of assuming in advance a particular constitutive relation, we
first approach the problem systematically by exploiting the implications  of
symmetry, and of the boundary condition of incipient failure at the free
surface (IFS).  We discuss, with reference to the  construction history of the
pile, the scaling ansatz of a {\em radial stress field} (RSF). This ansatz
transforms the partial differential equations into ordinary  differential
equations, which can be solved easily for all the closures considered later.
Using this, and the idea of {\em perfect memory} mentioned above, the range of
possible constitutive relations is greatly reduced.

After this,  we will solve
our equations for four specific closures in two dimensions; these are
incipient failure everywhere (IFE), Bouchaud-Cates-Claudin (BCC), fixed
principal axes (FPA), and finally the family of oriented stress-linearity (OSL)
closures, which  includes BCC and FPA as special cases.
All of these comply with our modelling strategy of seeking local constitutive
relations among stress components. For the
OSL model, stresses propagate along straight characteristics which can be
interpreted by analogy with wave propagation along ``light-rays". We thereby
arrive at a very simple geometrical picture of  stress
propagation, from which the forms of the stress profiles can be swiftly
deduced. For the OSL model, the effect of a small perturbation (adding a little
extra weight somewhere) is studied, and an appropriate Green function described.
This helps sharpen the idea of stress being carried along arches.


In Section \ref{sec3D} we
extend our calculations to the three dimensional conical sandpile. Because of
the larger number of stress components (some of which can be eliminated by
symmetry) a second constitutive equation is now required.
We study several possible choices for this secondary closure, and find that all
of these lead to qualitatively similar stress profiles. A comparison with the
experimental results of Smid  and Novosad \cite{smid} is then made. This shows,
firstly, that the RSF scaling assumption is well-verified, and, secondly, that
the data is fit rather well by the FPA model, without adjustable parameters.
Viewed alternatively as a comparison with the OSL model, which does have an
adjustable parameter (the tilt angle), the evidence suggests that parameter
values close to the FPA case must be chosen. The data thereby presents strong
evidence for the arching picture as an explanation of the dip.


Until the end of Section 3, we will have considered only the case where the
free surface of the pile is at the angle of repose. However, it is possible
experimentally to achieve piles which are flatter than this. We discuss this and
a number of related problems in Section 4, where, for simplicity, we restrict
attention to the FPA model in two dimensions. We show that it
matters how a sandpile is made: for example, if a flattened pile is created by
slicing wedges off the top of a steeper one, the stresses should differ from
those found by choosing a material with a lower repose angle to begin with. It
is in geometries such as this, that the dependence of the constitutive equation
on the construction history of the pile can be probed.

Section 5 contains a brief summary of our
approach and a concluding discussion.
Our calculations for the stress propagation in two-dimensional sandpiles,
using the OSL and FPA models, are new; as are our three dimensional results for
these models, and for BCC, although some of the FPA results were
outlined elsewhere \cite{nature}. For the IFE model, which is more
classical, the corresponding results may exist in the literature (though we
have not found them); in
any case we include them for  comparison.

\section{The two dimensional symmetrical sandpile}

\label{sec2D}

\subsection{Indeterminacy of stress continuity equation}


\label{secMCE}

As mentioned previously, the continuum approach to calculating the stress
distribution in a static sandpile immediately encounters an indeterminacy.
Indeed, the stress continuity equation in two dimensions reads (componentwise)
\begin{eqnarray}
\partial_r \srr + \partial_z \srz & = & 0 \label{eqpde}\\\nonumber
\partial_r \srz + \partial_z \szz & = & g
\end{eqnarray}
which, clearly, provides only two relations between the three independent
elements of the stress tensor \szz, \srr \ and $\srz=\szr$.

To ease the later generalization to three dimensions (Section 3) we here use
cylindrical polar coordinates, with $z$ measured {\em downward} from the apex
of the pile and $r$ a radial coordinate from the symmetry axis -- see
Fig.1(c). (In
two dimensions, $r= |x|$, with $x$ a cartesian coordinate.)
We have assumed that the granular medium has {\em constant density}, $\rho$,
thereby excluding segregation effects (see Section 1.2 above), and
have chosen units where $\rho=1$. The acceleration due to gravity
is denoted $g$; because it enters linearly, this could also be set to unity, but
we retain it for clarity. The stress tensor
$\sigma_{ij}$ is defined to be symmetric in $i,j$, as usual in the physics
literature.

In our cylindrical polar coordinate system the stress tensor is a function of
$z$ and $r$, where $r \ge 0$ by definition. However, in terms of cartesians
$(z,x)$, one would have both positive and negative $x$; in this
case, the normal stresses would be even functions of $x$ and the shear stress
an odd function \cite{BCC}.
(The latter holds because the unit vector along $r$ reverses sign at
the symmetry axis.) Confusion can be reduced by restricting attention to
the left half
of the pile (positive $x$) for which the two coordinate systems coincide. In any
case, on the symmetry axis itself ($r=0$), the shear stress must vanish by
symmetry, and the $z$ and $r$ directions are both principal axes there.

As well as the three stress
components
\srr, \szz \ and \szr \ , it will be useful to consider the
following three quantities: the average stress $P = (\srr+\szz)/2$,
the ``radius of Mohr's circle'' $R$ defined via
$R^2 = (\szz-\srr)^{2}/4 + \szr^2 $ \cite{Mohr},
and the (positive) angle of inclination $\Psi$ between
the $z$-axis to the major principal axis of the stress tensor (see
Fig.1(c)). In terms of these,
\begin{eqnarray}
\srr  & = & P-R \cos(2 \Psi) \label{eqPRPsitosigma}\\ \nonumber
\szz  & = & P+R \cos(2 \Psi) \\ \nonumber
\szr  & = & R \sin(2 \Psi)
\end{eqnarray}
Following the usage of the engineering literature \cite{nedderman}
we define a material point to be in an {\em active} state
if the normal stress \szz \ in the direction of the external compressive force
(here gravity) is larger that the stress
\srr \  perpendicular to it. On the symmetry axis ($r=0$), the $z$-axis is
the major principle axis and $\Psi=0$ in the case of an active state, whereas
for a passive state the major axis is $r$, and $\Psi=\pm\pi/2$.

The simplest model of a granular medium is known as the
{\em ideal cohesionless Coulomb material}.
The Coulomb model plays the same role in the
study of granular materials as the Newtonian fluid does in
viscous flow, and we will use it here.
Plastic failure occurs in a given material element if
there exist a plane
defined by a unit normal $\bf n$ (or angle of inclination $\tau =
\sin^{-1}({\bf n.\hat z}$)) through this element, on which the shear forces \sns
\ exceeds  a given fraction of the normal force \snn \
across the plane \cite{coulomb}. Conversely the element is stable if, for all
such planes,
\begin{equation}
|\sns | \leq  \tan(\phi) \snn
\label{eqCoulomb1}
\end{equation}
For a material with {\em cohesion}, a constant $c$ is added to the right
hand side; we treat only the cohesionless case.
The Coulomb yield criterion
can then alternatively be expressed as
\begin{equation}
\Yield \equiv  {R\over P\sin(\phi)} \leq 1
\label{eqCoulomb2}
\end{equation}
(Put differently, ``the yield locus must not cut Mohr's circle" \cite{Mohr}.)
The coefficient $\tan(\phi)$ in eqn.(\ref{eqCoulomb1}) is the coefficient of
static friction of the material; elementary arguments show that $\phi$ is then
the angle of repose \cite{nedderman} (defined, as usual, as the inclination of
the free surface to the
horizontal; see Fig.1(c)).

\subsection{IFS boundary conditions}

We now use the yield criterion to specify the stresses
on the surface of the pile. In doing this, we
neglect the small Bagnold hysteresis angle (as mentioned in Sec. 1.1)
and demand that the {surface} of a pile, constructed from a point source and at
its angle of repose, is in a state of incipient slip. (Sandpiles constructed
differently, for which this is not the case, are considered in  Section 4).

First we note that (in two dimensions) all stress components have to
vanish on the surface:
\begin{equation}
\srr(S=1)=\szz(S=1)=\szr(S= 1)=0
\label{eqby1}
\end{equation}
Here we have introduced, for reasons that will be clarified later, a scaling
variable $S=r/(cz)$ with
$c = \cot(\phi)$. (Hence the equation of the free surface is $r=cz$, or $S=1$.)
The vanishing of the stresses is a direct consequence of the yield criterion, as
we now show by considering the stress components in a rotated
coordinate system $(n,m)$ (see~Fig.1(c)).
For a system inclined at angle $\tau$ to the vertical, one has
\begin{eqnarray}
\snn & = & \cos^2(\tau) \srr + \sin^2(\tau) \szz -
2 \sin(\tau) \cos(\tau) \szr \label{eqrot}\\ \nonumber
\sss & = & \sin^2(\tau) \srr + \cos^2(\tau) \szz +
2 \sin(\tau) \cos(\tau) \szr \\ \nonumber
\sns & = & - \sin(\tau) \cos(\tau) (\szz-\srr) +
           (\cos^2(\tau) - \sin^2(\tau)) \szr
\end{eqnarray}
Now choosing $\tau = \pi/2-\phi$, so that $n$ is normal to the surface,
we require that the normal stress \snn \ at the free surface has to vanish
(this assumes that no external forces act there). The yield
criterion eqn.~(\ref{eqCoulomb2}) then requires
\begin{equation}
\sss^2 \cos^2(\phi) + 4 \sns^2 \leq 0 \label{another}
\end{equation}
Accordingly, the remaining two stress components \sss \ and \sns \ must
also vanish, and the stress tensor is zero in the $(n,m)$,  and hence in the
$(r,z)$, coordinate system.

The criterion that the surface of the pile is a slip plane, not only
implies that
the stresses on the surface vanish, but also fixes their ratios
in its immediate neighbourhood.   Demanding equality in
eqn.~(\ref{eqCoulomb1}) as the surface is approached, we obtain the
condition
\begin{equation} \hbox{\rm lim}_{S\to 1}{\sns(S)\over \snn(S)} =-\tan(\phi)
\label{eqby2a}
\end{equation}
(the sign can be confirmed from eqn. (\ref{eqrot})).
Applying also eqn.(\ref{another}) (with equality) in this limit gives a second
condition:
\begin{equation}
\hbox{\rm lim}_{S\to 1} {\sss(S)\over\snn(S)}=1 +
2\tan^2(\phi)\equiv 1/\eta_0
\label{eqby2b}
\end{equation}
where the final notation will prove convenient later.
By rotating using  eqn.~(\ref{eqrot})
these can be written in the $(r,z)$-system as
\begin{eqnarray}
\hbox{\rm lim}_{S\to 1}{\szr(S)\over \srr (S)} & = & \tan(\phi)
\label{barmy}\\ \nonumber
\hbox{\rm lim}_{S\to 1}{\srr(S)\over \szz (S)} & = & \eta_0
\end{eqnarray}
(The results in $(m,n)$ and in $(r,z)$ coordinates look rather
similar because, as it turns out, the two
frames are related by a reflection through the major principal axis
\cite{bisect}.)

The requirements expressed by
equations (\ref{eqby1}, \ref{barmy}),
represent a set of ``boundary conditions" which we denote IFS (incipient failure
at the surface).  Along with the stress continuity equation, these form the
boundary value problem for determining the stress profile of a sandpile.
At first sight there may appear to be more boundary conditions than required;
however (as emphasised before) to close the problem in two dimensions,
we will need a constitutive relation between stress components, which is yet
to be chosen. One can therefore view any extra ``boundary
conditions" as constraints limiting this choice. We next develop
some  general scaling arguments which, combined with some other physically
motivated simplifications, further restrict the choice of constitutive
equation.

\subsection{Scaling analysis}

\label{secScaling}

The basis of our scaling approach is to assume that the macroscopic
material properties of a granular medium (under gravity) are independent of
length scale. Obviously,
any such medium has a characteristic length associated with the grain size, but
in a continuum description this should not be important. The scaling hypothesis
supposes that not only this length, but also any other characteristic
length-scale that the medium possesses, is either extremely small, or else
extremely large, compared to the scales probed in a macroscopic sandpile
experiment. As mentioned previously, and shown below, our scaling
assumption is
well-verified in the experiments of Ref.\cite{smid}. However, those experiments
are on piles of a limited size range (20 to 60 cm high). It is possible that for
smaller or larger piles our scaling assumptions would break down, due to a
characteristic length arising from size segregation \cite{liffmann}, for
example.
A corollary of our scaling assumption is that no relevant intrinsic
scale exists for stresses: otherwise, this scale could be compared with the
gravitational stress to give a length. Thus we exclude, for example,
deformable particles whose elastic modulus sets a stress scale, and thereby
a ``deformation length" (which for rigid particles is infinite).
These simpifications, though guided by experiment, are not physically obvious
{\em a priori}; the problem deserves more careful experimental study to
determine
the limits to the scaling regime for real granular materials.

Assuming that the medium indeed has no intrinsic characteristic length, the
stress distributions in all piles formed the same way (of the same material)
should be similar. Hence we search for a scaling solution of eqn.~(\ref{eqpde})
of the form:
\begin{equation}
 \sigma_{ij} = g z s_{ij}(S)
\label{eqonlyH}
\end{equation}
with the scaling variable $S=r/cz$ was introduced previously. In anticipation
of this ansatz, our earlier discussion of the boundary conditions was couched in
terms of $S$.  These boundary conditions impose $s_{ij}(1)=0$, and also fix
ratios of derivatives of
$s_{ij}$ (see eqns.(\ref{eqby1},\ref{barmy}) above). The functions $s_{ij}$
have to be
continuous everywhere; however, we are dealing with hyperbolic equations, and
the stresses need not be differentiable
\cite{discfoot}.

The scaling ansatz, eqn.~(\ref{eqonlyH}),
reduces the partial differential equations for stress continuity,
eqns. (\ref{eqpde}), to the following
ordinary differential
equations:
\begin{eqnarray}
\grr'/c + \gzr - S \gzr' & = & 0 \label{eqDE}\\ \nonumber
\gzr'/c + \gzz - S \gzz' & = & 1
\end{eqnarray}
The primes denote derivatives with respect to the scaling variable $S$;
recall that $c=\cot(\phi)$.
Solutions of eqn.~(\ref{eqDE}) are usually called
``{\it radial stress fields}'' \cite{radialstressfield}
and we refer to (\ref{eqonlyH}) as RSF scaling.
>From the scaling behaviour of $s_{ij}$ follows a corresponding scaling of the
mean stress,
$P(r,z) = g  z
\tilde{P}(S)$, the radius of Mohr's circle, $R(r,z) = g z \tilde{R}(S)$, and the
angle of the major principle axis $\Psi(r,z)={\Psi}(S)$.

We now ask, what are reasonable closure relations
consistent with the radial stress field form of eqn.~(\ref{eqonlyH})?
On physical grounds we first impose the requirement of {\em locality}:
the unknown stress component in a material element depends on the known ones
in that element, and not elsewhere. (As mentioned previously, this would fail
if distant loads were able to rearrange the grains themselves in a given
neighbourhood.) The
most general form consistent with our scaling ansatz is then
\begin{equation}
 \srr/\szz = \Closure(\szr/\szz,S)
\label{eqnointrinsic}
\end{equation}
where the absence of an intrinsic stress scale, noted above, means that only
{dimensionless ratios} of stresses can enter.

To restrict \Closure \ further, we now invoke the assumption of {\em perfect
memory} mentioned in the introduction: that the constitutive relation in a
material element is determined at the time of its burial and is not
subsequently
altered. Now, it is clear that each material element is buried while
just at the surface of the pile ($S=1$), after which it experiences
continually decreasing values of $S$ as the pile gets larger. Hence any
explicit dependence of \Closure \ on  $S$ would violate the
perfect memory assumption. Accordingly, we must have
\begin{equation}
\srr/\szz = \Closure(U)
\label{eqmemory}
\end{equation}
where we have defined the reduced shear stress
\begin{equation}
U(S)=\szr/\szz=\gzr/\gzz
\label{udef}
\end{equation}

An exception to the above argument should be made for
material elements on the central axis, which are buried, and remain forever,
with a value
$S=0$.  In two dimensions the centreline divides grains which have rolled
to the left
from those which have rolled to the right, and which therefore have had
qualitatively different histories \cite{leftright}. Accordingly, there is no
requirement that
\Closure \  behave smoothly at the origin; and although the constitutive models
studied below all appear analytic when expressed in polar ($z,r$) coordinates,
for some of them \Closure \ does become singular
on the symmetry axis, when a cartesian ($z,x$) coordinate system is employed.
(The models with this property are the OSL models, including FPA, but with the
exception of BCC.)

We can now reformulate the radial stress field
eqns.~(\ref{eqDE}) as
\begin{eqnarray}
\frac{d\gzz}{dS} & = & \frac{\ccc \eee -\bbb \fff}{\bbb \ddd - \aaa \eee}
\label{eqRK}\\ \nonumber
\frac{d\gzr}{dS} & = & \frac{\aaa \fff-\ccc \ddd}{\bbb \ddd - \aaa \eee}
\end{eqnarray}
Here we have introduced the notations
\begin{eqnarray}
\aaa & = & -\frac{1}{c} \frac{U^2 d(\Closure(U)/U)}{dU} \label{eqabcdef}\\
\nonumber
\bbb & = & -S + \frac{1}{c} \frac{d\Closure(U)}{dU} \\ \nonumber
\ccc & = & \gzr(S) \\ \nonumber
\ddd & = & -S \\ \nonumber
\eee & = & 1/c \\ \nonumber
\fff & = & \gzz(S) -1
\end{eqnarray}
where only the first two functions explicitly involve the closure
relation. Similarly we can also reformulate the Coulomb yield criterion,
eqn. (\ref{eqCoulomb2}) as:
\begin{eqnarray}
1 \geq \Yield(U)^2   =
\frac{(1-\Closure(U))^{2}+4 U^{2}}{(1+\Closure(U))^{2} \sin^2(\phi)}
\label{eqCoulomb3}
\end{eqnarray}
eqns.~(\ref{eqRK}) give a systematic procedure
for solving (at least numerically) the boundary problem for any specified
constitutive relation \Closure ; eqn.~(\ref{eqCoulomb3}) then allows one to
check
its stability. The latter step is necessary to ensure that the
yield criterion (marginally satisfied at the surface) is not violated
deep inside the pile; any closure for which such violations arise must be
rejected.

\subsection{Asymptotic behaviours}

\label{secSM}

Without specifying \Closure \ further, we can now
use the scaled continuity equations, eqn.~(\ref{eqDE}),
to examine the possible asymptotic behaviours close to the free surface, and
close to the symmetry axis of the pile.

As mentioned previously, all the stresses vanish at the free surface ($S \to
1$). Asymptotically we expect them to vanish as power laws
$\grr = a_1 (1-S)^\alpha$, $\gzz = b_1 (1-S)^\beta$,
$\gzr = d_1 (1-S)^\delta$
($a_1,b_1,d_1,\alpha,\beta,\delta > 0$).
The IFS condition, eqn.~(\ref{barmy}), requires that \grr \ and \gzz \ have
to vanish with the same power,
$\alpha=\beta$, since their ratio approaches a constant.
More generally one finds by substituting the power law forms into
eqn.~(\ref{eqDE}) that the stresses
on the surface have to vanish {\em linearly} : $\alpha=\beta=\delta=1$.
This applies for any
choice of the closure \Closure .
(For the models solved below, this linear behaviour is
visible in
Fig.2(a).)
Only one of the coefficients $a_1,b_1,d_1$
then remains free; using (\ref{eqDE}) we find
$d_1 = a_1/c$ and
$-d_1/c +b_1 =  1$.
Using again the IFS boundary condition,
eqn.~(\ref{barmy}), we obtain finally \cite{overdetermined}
\begin{eqnarray}
a_1 & = & 1/(1 + \tan^2(\phi)) \label{eqabdonslope}   \\ \nonumber
b_1 & = & (1+ 2\tan^2(\phi))/(1 + \tan^2(\phi)) \\ \nonumber
d_1 & = & \tan(\phi)/(1+\tan^2(\phi))
\end{eqnarray}
This completes the specification of the
asymptotic behaviour near the free surface.
The average reduced stress on the surface is $\tilde P=1-S$ and
the reduced Mohr's radius $\tilde R=\sin(\phi) (1-S)$.
Since, from eqn. (\ref{eqPRPsitosigma}),
$\cot(2\Psi)=\left| (b_1-a_1)/2d_1 \right|$ we can solve
for the direction of the major principal axis at the free surface
\begin{equation}
\Psi(1) = \psi \equiv (\pi-2\phi)/4
\label{eqPsionslope}
\end{equation}
This asymptote for $\Psi(S\to1)$ bisects the angle between the vertical and the
free surface itself.

A similar analysis can be made of the stresses close to the symmetry axis of the
pile ($S \to 0$). Again without knowing details of \Closure , we can establish
for $S\to 0$ a
solution involving a linearly vanishing shear stress
$\gzr=d_0 S$,
a vertical normal stress of the form $\gzz = b_0 + b_{cusp} S$, and
a flat horizontal normal stress  $\grr = \eta b_0$.
Here $\eta \equiv \Closure(0)$
is the ratio of horizontal and vertical stresses in the middle of the pile.
One can also show that $b_0=1-d_0/c$, with $b_0, d_0$ positive. The existence of
a cusp,
$b_{cusp}
\neq 0$, is associated with a breakdown in the smoothness of the
closure relation \Closure \ on the symmetry axis; as discussed above, this is
physically permissible, and is a distinguishing feature of the new models
introduced in this paper.

The results of this and the preceding section were obtained by combining RSF
scaling with the perfect memory assumption, without further restriction on
\Closure . In the next few sections we finally consider various model
constitutive relations with which to close the equations and thereby calculate
explicit stress profiles for the sandpile.

\subsection{The IFE model}

\label{secIFE}

The traditional \cite{drescher,nedderman,sokolovskii} assumption of
Incipient Failure
Everywhere (IFE) means that the granular material is everywhere marginally
unstable;  the frictional forces are fully mobilized and $\Yield = 1$.
(Accordingly the two principal stresses are everywhere in fixed ratio.)
Indeed, we can solve eqn.~(\ref{eqCoulomb3}) with equality to find:
\begin{equation}
\Closure(U) = \frac{1}{\cos^{2}(\phi)}
\left( (\sin^{2}(\phi)+1) \pm 2 \sin(\phi) \sqrt{1- (\cot(\phi) U)^2} \right)
\label{eqgrrIFE}
\end{equation}
where $U(S)=\gzr/\gzz$ is the reduced shear stress introduced previously.
Here the negative sign must be chosen: this corresponds to requiring downward
(rather than upward!) incipient slip of the grains at the free surface.
The resulting IFE constitutive equation then fixes the two
functions $\aaa(S),\bbb(S)$ defined earlier, as follows:
\begin{eqnarray}
\aaa & = &
\frac{1}{\cos^2(\phi)}  \left(1+\sin^2(\phi)-2 \sin(\phi)
\sqrt{1-(\cot(\phi)U)^2} \right) \label{eqabIFE}\\ \nonumber
\bbb & = & -s +  \frac{2 U}{c \sin(\phi) \sqrt{1-(\cot(\phi)U)^2}}
\end{eqnarray}
thereby closing the the radial stress field
equations~(\ref{eqRK}). We have not obtained an analytical
solution for this model, but a numerical solution was readily found by a
standard Runge-Kutta procedure (by shooting from the middle of the pile to the
boundary conditions on the surface). The reduced stresses \gzz, \grr \
and \gzr \ are shown for the case $\phi=30^o$
in Fig.2(a).
(Also shown are results from the other
models discussed below.)
The IFE model give a smooth ``hump" in $s_{zz}$.

\subsection{The BCC model}

\label{secBCC}

In place of the IFE assumption that the {\em principal} stresses
are proportional, the BCC model \cite{BCC} assumes instead the
proportionality of
vertical and horizontal normal stress components:
\begin{equation}
{\sigma_{rr}\over \sigma_{zz}}\equiv \Closure(U) = \eta
\label{bccclos}
\end{equation}
This assumption, which is perhaps the simplest possible, is related (but not
identical) to one made in the classical work of Janssen
\cite{janssen,nedderman,drescher}. Invoking also the IFS boundary
conditions, complete
results were obtained analytically for the two dimensional sandpile \cite{BCC}.
These results for the stresses, the yield function $\Yield$ and the
orientation angle $\Psi$ of the major principal axis are shown (alongside
those of IFE and FPA) in
Figs.2(a-c).

The IFS condition (eqs.\ref{eqby1},\ref{eqby2a},\ref{eqby2b}) in fact
requires $\eta=\eta_0$ (where $\eta_0$ is defined in eqn.\ref{eqby2b}).
The stresses are piecewise linear with
a cusp at $S = S_0=\CO/c$, where
$\CO = \sqrt\eta_0$ and, as shown by BCC, $\CO/c$ is strictly less than unity.
The inequalities $\CO/c \le S \le
1$ define an ``outer region" of the pile in which the stresses obey $s_{zz}  =
(1-S)/(1 - (\CO/c)^2)$ and $s_{zr}  =  (1-S)\CO^2 c/(c^2 -\CO^2)$ . These match
at
$S=S_0$ onto an ``inner region" ($0
\le S \le \CO/c$), in which there is a plateau for the vertical normal stress,
$s_{zz} = 1/(1 + \CO/c)$, while the shear stress vanishes linearly on the
central axis,  $s_{zr} =
S \CO/(1+\CO/c)$.
As shown in eqn.(\ref{eqPsionslope}) above, the inclination angle of the major
axis obeys at the surface $\Psi(1) = \psi$. For the BCC closure in two
dimensions, this value is maintained throughout the outer region, while in the
inner region $\Psi(S)$ obeys $\tan(2\Psi) = 2 \CO S/(\CO^2 -1)$. Hence $\Psi(S)$
vanishes smoothly at $S=0$.

\subsection{The Fixed Principal Axis (FPA) model}

\label{secFPA}

Neither the IFE nor the BCC closure gives a``dip" in two dimensions (nor
in three as shown below).
We therefore propose a new hypothesis \cite{nature} which appears to
capture, within a
fully consistent continuum theory, the physics of arching (as expounded by
Edwards and coworkers \cite{EO,EM}).
Specifically, we postulate that the major principal axis of the stress
tensor has a {\em fixed angle of inclination} to the
downward vertical: $\Psi(S)$ is constant. We first describe the results
and afterwards discuss in more detail the physical content of this model.

The FPA hypothesis provides a local constitutive equation by assuming
that the principal stress axes in a material element have constant orientation
fixed at the time of its burial. However, with the exception of those lying
right on
the symmetry axis, all such elements were first buried at the surface of
the pile ($S=1$).
Since the IFS boundary condition already fixes
$\Psi(1) = \psi \equiv (\pi-2\phi)/4$, our FPA model requires
\begin{equation}
\Psi = \psi
\end{equation}
everywhere.
Using the results of Sec. 2.1, one finds immediately that this is equivalent to
the following constitutive relation:
\begin{equation} {\srr\over \szz} \equiv
\Closure (U) =  1 -2\tan(\phi)\, U \label{fixedpa}
\end{equation}
Note that if $r$ is replaced by $x$ (cartesian coordinates) this
becomes
\begin{equation} {s_{xx}\over \szz} \equiv
\Closure (U) =  1 -2\,\sign(x)\tan(\phi)\, U \label{fixedpaxz}
\end{equation}
The $\sign(x)$ factor is a reminder that, in the FPA model (unlike BCC)
our $\Closure (U)$ is nonanalytic at the symmetry axis of the pile. A
compact
way to write eqn. (\ref{fixedpaxz}) is \cite{nature}
$\Closure (U) =  1 -2\tan(\phi)\, |U|$,
though (see Section 4 below), this version is not equivalent for all
construction histories.

Since the repose angle
$\phi$ (and thereby
$\psi$) is a material parameter fixed by experiment, the FPA model gives a
complete closure of the
$d=2$ sandpile problem. The resulting equations are linear.
Their structure is clearest when written in terms of the {unscaled} stress
components
$\sigma_{ij}$;
substituting eqn.(\ref{fixedpa}) and
eqn.(\ref{udef}) into eqn.(\ref{eqpde}) (taking $\sigma_{zz}$ and
$\sigma_{rz}$ as
the independent variables) gives:
\begin{eqnarray}
\partial_r \srz + \partial_z \szz & = & g \\
\partial_r (\szz -2\tan\phi \,\srz) + \partial_z \srz & = & 0
\end{eqnarray}
which can be rewritten
\begin{equation}
(\partial_z-c_1\partial_r) (\partial_z-c_2\partial_r)\sigma_{ij} = 0
\label{FPAwave}
\end{equation}
with
\begin{eqnarray}
c_1+c_2 & = & -2\tan(\phi) \\
c_1c_2 & = & -1
\end{eqnarray}
where we $c_1$ ($c_2$) to denote the positive (negative) roots. A little
manipulation then shows that $c_1 = \tan(\psi)$ and
$c_2 = -\cot(\psi)$.
A similar equation for the stresses, but with $c_1=-c_2 = \sqrt{\eta_0}$, was
obtained for the BCC model \cite{BCC}, in which case eqn.(\ref{FPAwave}) becomes
the wave equation in two dimensions. We shall call (\ref{FPAwave}) a ``wave
equation" even when $c_1+c_2\neq 0$; under these conditions, it becomes an
ordinary wave equation (with equal velocities) if tilted coordinate axes are
chosen
\cite{tilt}.

A complete solution is readily found and is given explicitly (in the
context of the more general OSL model) in the next Section. As with BCC, one
finds  for $s_{zz}(S)$ a piecewise linear function, with inner and outer
regions.
The material in the outer region again saturates the yield criterion
(\ref{eqCoulomb3}) whereas the inner part does not;
these regions are separated by a cusp at $S=S_0 = c_1/c$. For the FPA model,
there is always a dip in $s_{zz}$ at the centre of the pile.  The dip takes the
form of a cusp at $S=0$ and is connected with
the nonanalyticity of $C(U)$, which
reflects the sudden change in the direction of the major principal axis on
passing through the central axis of the pile. The maximum vertical stress (at
$S=S_0$) is a factor $(1+2\tan^2(\phi))$ times larger than the value at
$S=0$ (the latter is always finite). These results are compared with the BCC
and IFE models in Fig.2(a-c).

At first sight the requirement of fixed orientation of the stress tensor
($\Psi=\psi$) is at odds with the fact that, on the centre line of the pile,
there are no shear stresses and so the horizontal and vertical directions
must be
principal axes ($\Psi=0$). This paradox can be
resolved by noting that on the centre line the stress in the FPA model is
actually  isotropic, thus satisfying both criteria at once.

The correspondence between the FPA model and the Edwards arching picture becomes
clearer on considering the characteristics of the wave equation,
(\ref{FPAwave}), which are straight lines of slope $c_1$ and $c_2$ (as discussed
further in Section 2.9 below), Fig.3. Since for the FPA model  $c_1 =
\tan(\psi)$ and
$c_2 = -\cot(\psi)$, the characteristics are at rightangles to one another;
moreover, {\em they coincide with the principal axes of the stress tensor}.
It is this special property of the FPA model that we believe embodies
Edwards' physical picture of arches \cite{EO,EM}. The stress arising from the
weight of an element of sand propagates along two straight characteristics, one
at
$\psi$ to the vertical (which we identify as the ``arch  direction", coincident
with the major principal axis) and the other at rightangles.
As shown in Section 2.9 below, the {\it majority} of the stress is carried
by the
outward characteristic (slope $c_1$).
The material can
therefore be viewed as a set of nested arches (Fig.1(b)) down which most of the
stress propagates. (This ties in with Edwards' idea of ``lines of force"
\cite{EM}.) However, {\em a minority} of the stress is transmitted instead from
one arch to its inner neighbour; this transfer imparts mechanical stability to
the outer, incomplete, arches. Since the principal axes and the
characteristics coincide, there are no shear forces acting at the interface
between successive arches, which are therefore effectively in {\em frictionless}
contact with one another. This seems to be as close as one can get, within a
consistent continuum theory, to the intuitive picture of arches as
{independent} load-bearing structures.

We emphasize that for a sandpile constructed from a point source, the FPA
model can be
viewed in two ways; either as a {\em macroscopic} hypothesis concerning the
transmission of stresses at the scale of a pile (principal axes fixed in space),
or as a
{\em microscopic} hypothesis concerning the way the growth history of
the pile is locally encoded. Our own
modelling approach, based on local constitutive relations among stresses,
corresponds to the latter view. Indeed,
after
making the assumption of perfect memory, we have to choose one scale-free
property (the
constitutive equation) to be ``remembered" by any element from the moment of its
burial: and the choice made by FPA corresponds to remembering the orientation of
the stress tensor (principal axes fixed in time). Note that, once the basic FPA
assumption is made, there is no free parameter left in the theory (at least, not
in two dimensions), since
$\phi$ is fixed by experiment.

It seem plausible that the construction of the pile (by a series of
avalanches at
the surface) imparts to the local packing of grains  a permanent sense of
direction. If so, the FPA constitutive relation is perhaps the simplest model
for how this ``orientiational memory" within an element could determine
the constitutive relation among stresses arising there subsequently.
A possible (though not a necessary) interpretation of this directional
memory is in terms of a fabric tensor $\lambda_{ij}$
\cite{fabric}. For example, one could postulate that $\lambda_{ij}$ was constant
throughout the medium (when expressed in cylindrical polar coordinates)
and moreover had a principal axis $\Psi$ bisecting the free surface and the
vertical. If this were true, the FPA constitutive relation would reduce
to the commutation requirement $\sigma_{ij}\lambda_{jk} =
\lambda_{ij}\sigma_{jk}$.

Apart from its appealing simplicity, however, we have no detailed mechanistic
justification for the FPA model in terms of the fabric tensor or any similar
quantity: why should the orientation of the principal
axes be remembered, rather than something else? (For example, in the IFE model
each element ``remembers" instead that it was at the critical threshold for slip
when buried, and remains so forever after.)
We therefore propose the FPA
model as a phenomenological hypothesis to be tested against experiment.  It is
interesting, in that context, to consider alternative closures; we do this next,
by embedding the FPA model within a broader scheme.

\subsection{The Oriented Stress Linearity (OSL) model}

\label{secOSL}

A sandpile is formed by layerwise deposition of particles
that have rolled down its free surface. Thus the grains of sand may
end up arranged in a packing that locally distinguishes the directions
toward and away from the central axis.
An arching effect can arise if this anisotropy tends to direct
stresses outward from the centre, thus
``screening" the central part of the pile from the
added load of new layers. This offers a possible way to explain the dip;
and indeed the FPA model can be viewed in exactly these terms. However,
it is not unique in this respect.

A more general approach can be generated by an adaptation
of the BCC model, in which it was assumed that $\sigma_{rr} = \eta\sigma_{zz}$.
BCC thus singles out for special treatment the vertical and horizontal
normal stresses. We now define the oriented stress linearity (OSL) model by
assuming a similar linear relationship between normal stresses, not in a ($z,r$)
coordinate system, but in a tilted one ($n,m$). The latter system is
now characterized by an {\it arbitrary} (but constant) tilt angle
$\tau$ to the vertical, and is related to ($z,r$) via the transformation
equation (\ref{eqrot}). (In general this $n,m$ system does {\em not} coincide
with the one used earlier to discuss the IFS boundary condition.)
In the tilted coordinates, we now require (following BCC)
that the two normal stresses, \snn \ and \sss \
are proportional:
\begin{equation}
\snn = K \sss
\end{equation}
(see Fig.1(c)).
Despite its formal similarity,  the OSL closure differs
critically from BCC in that it violates the assumption, tacitly made by
BCC, that
the properties of the medium vary smoothly as one passes through the centre line
of the pile; it thereby allows a cusp (dip or hump) to arise in $s_{zz}$.

Leaving the angle $\tau$ and the constant $K$ free for the moment,
we use the rotation eqn.~(\ref{eqrot})
to obtain, in ($z,r$) coordinates, the OSL constitutive relation
\begin{equation}
{\sigma_{rr}\over\sigma_{zz}} =\Closure(U) = \eta + \mu \, U
\label{eqlinclosure}
\end{equation}
(where, as always, $U = \sigma_{rz}/\sigma_{zz}$). As with FPA, in
cartesians ($z,x$) this becomes
\begin{equation}
{\sigma_{xx}\over\sigma_{zz}} =\Closure(U) = \eta + \mu\, \sign(x)\, U
\label{eqlinclosurexz}
\end{equation}
whereby the singularity on the centreline becomes apparent.

The constants $\eta$ and
$\mu$ obey:
\begin{eqnarray}
\eta & = & \frac{K-\tan^2(\tau)}{1-K \tan^2(\tau)} \label{eqetamuKtau}\\
\nonumber
\mu & = &  \frac{2 (K+1) \tan(\tau)}{1-K \tan^2(\tau)}.
\end{eqnarray}
Clearly the BCC model
corresponds to $\eta = \eta_0(\phi)$ (defined in eqn.\ref{eqby2b}) and $\mu=0$,
whereas the FPA model, eqn.~(\ref{fixedpa}), is obtained by setting $\eta = 1$,
$\mu = -2\tan(\phi)$. Both are thereby special cases of OSL \cite{ifebcclink}.
Note that pairs of OSL coordinate systems inclined through angles $\tau$ and
$\tau + \pi/2$ (or $\tau-\pi/2$) are identical, subject to interchanging the $m$
and $n$ axes; they give the same values of $\eta$ and $\mu$ in
(\ref{eqlinclosure}) and hence the same stress profiles in the pile.

The coefficients $\eta$ and $\mu$ (or equivalently $K$ and $\tau$) in
the OSL model are not independent: an equation between them can be found
from the
IFS boundary condition as formulated in eqns.~(\ref{barmy}). For a given
repose angle $\phi$, this condition restricts the OSL parameters to the ``IFS
line" :
\begin{equation}
\eta= \eta_0 (1 - \mu \tan(\phi))
\label{eqstabline}
\end{equation}
Hence the OSL model has
one remaining free parameter (unlike BCC or FPA, which have none). The IFS lines
in the
$(\mu,\eta)$-plane are shown for two values of the friction angle $\phi$
in Fig.4.

The OSL constitutive relation eqn.(\ref{eqlinclosure}) can be substituted
into the stress continuity equations (\ref{eqpde}) to give
\begin{eqnarray}
\partial_r \srz + \partial_z \szz & = & g \label{oslform}\\ \nonumber
\partial_r (\eta\szz +\mu \srz) + \partial_z \srz & = & 0
\end{eqnarray}
from which we can obtain a wave equation
of the form (\ref{FPAwave}), as discussed already in the context of the FPA
model. In this more general case, however, $c_1$ and $c_2$ are the positive
and negative roots respectively of
\begin{equation}
c_{1,2} = \frac{1}{2}(\mu \pm \sqrt{\mu^2 + 4 \eta})
\label{eqvelocities}
\end{equation}
The propagation velocities become equal in magnitude if coordinate
axes are rotated by the tilt angle $\tau$.

The resulting stress propagation equations
can be solved without difficulty.
As with the FPA model, there are inner and outer regions which meet at $S
=S_0=c_1/c$;
in the outer region we obtain
\begin{eqnarray}
\gzz & = & s_{*} (c-\mu) (1-S) \label{eqoutside}\\ \nonumber
\grr & = & s_{*} \eta c (1-S)\\ \nonumber
\gzr & = & s_{*} \eta (1-S)
\end{eqnarray}
where we have introduced the constant
\begin{equation}
s_{*} = \frac{c}{c^2-\mu c -\eta} = \frac{c c_1}{(c c_1 + \eta)(c -c_1)}
\end{equation}
In the inner region ($0 \le S \leq c_1/c$) we find
\begin{eqnarray}
\gzz & = & s_{*} (c-c_1)/c \  (c_1 - \mu S) \label{eqinside}\\ \nonumber
\grr & = & s_{*} \eta c_1 (c-c_1)/c \\ \nonumber
\gzr & = & s_{*} \eta (c-c_1)/c \  S
\end{eqnarray}
As stated already for FPA and BCC, we thus obtain stress profiles that are
piecewise linear functions of $S$. We see from (\ref{eqinside}) that a dip in
$s_{zz}$ is present so long as $\mu < 0$. This applies for OSL models on
the part
of the IFS line which lies in the left hand half plane, Fig.4; such models
are separated by the BCC model ($\mu = 0$) from OSL models with a hump
($\mu > 0$).

As described earlier in connection with eqn.~(\ref{eqCoulomb3}), a further
check on the consistency of the model should now in general be made: we require
that the yield threshold is not exceeded within the
pile. The above equations show that the threshold is exactly saturated, not
only at the surface (IFS) but throughout the outer region. However, there is
also the possibility of yield in the inner region; when this happens, it first
occurs at the very centre of the pile \cite{midslip}. In this
neighbourhood, where shear stresses are negligible,
the Coulomb criterion eqn.~(\ref{eqCoulomb3})
simplifies to
\begin{eqnarray}
\eta_{min} \leq \eta \leq \eta_{max} \label{limits}\\
\nonumber
\eta_{min} = {1-\sin\phi\over1+\sin\phi} = \eta_{max}^{-1}
\end{eqnarray}
Where $\eta_{min}$ is known as``Rankine's coefficient of active earth
pressure" \cite{nedderman}.
Thus, for a given repose angle $\phi$, acceptable OSL parameters lie on the the
segment of the IFS line bounded by eqn.(\ref{limits}) (dash-dotted lines in
Fig.4). Outside this range, there is either too deep a dip or too
high a hump, leading respectively to passive or active failure of the material
at the centre of the pile.

The FPA model lies in the ($\eta,\mu$) plane at the point where
the IFS line crosses $\eta = 1$. It divides those OSL models which, on the
central axis of the pile, have active behaviour ($\Psi(0)=0$), from those which
are passive there ($\Psi(0)=\pm\pi/2$).  For the OSL models generally, the
orientation of the principal axes varies smoothly as one passes from left to
right through the centre of the pile (though the constitutive equation is
nonanalytic there). The sole exception to this is FPA,  which has instead a
discontinuity in $\Psi$ at $S=0$, for the reasons discussed in
the previous section.  This highlights the fact that FPA is the only model in
the OSL family for which the geometrical picture of ``nested arches" (Fig.1(b)
and Sec. 2.6 above) can definitely be said to apply.

\subsection{Linear models and ``light-rays"}

\label{secLinear}

We have seen that in the OSL model, the stresses propagate with a wave
equation in which
$c_1$ and
$c_2$ are the positive and negative roots of eqn.(\ref{eqvelocities}).
The characteristics of
this hyperbolic equation are thus straight lines of slopes $c_1$ and $c_2$. This
means that, if a perturbation is made at some point in the pile (for example,
increasing the weight of a certain element of sand), the resulting information
travels along two ``light rays" (together called a ``light cone" in
Ref.\cite{BCC}). Since the stress propagation equations are linear, the entire
stress distribution can be constructed by summing the contributions from all
elements of sand propatated along suitable rays; this offers an instructive
geometric insight into the problem.

First we consider the Green function which describes the stress perturbation
arising from a point source of weight. Such a source term violates the
left-right (``cylindrical") symmetry of our
two-dimensional system; to deal with it we must introduce cartesians ($z,x$) as
opposed to the polar coordinates ($z,r$) used so far. In such coordinates,
eqn.~(\ref{FPAwave}) is virtually unchanged:
\begin{equation}
(\partial_z\pm c_1\partial_x) (\partial_z\pm c_2\partial_x)\sigma_{ij} = 0
\label{greenwave}
\end{equation}
where the $+$ signs apply for $x>0$ and $-$ for $x<0$.
Our source term then
consists of adding $\Delta g(z,x) =\delta(x-x_0)\delta(z-z_0)$ to the right
hand side of the second member of the stress continuity equation
(\ref{eqpde}), in which $x$ now replaces $r$.
This yields an
inhomogeneous form of (\ref{greenwave}) with derivatives of the
delta-function on
the right hand side. The algebraic form of the Green function
is complicated (we do not write it out explicitly here) but its geometric
interpretation is
relatively simple, as shown in Fig.5(a).

Of the stress $\sigma_{zz}$ contributed by a small element of sand, a fraction
$A_1$ propagates along the outward light ray and $A_2$ along the inner ray.
(Note that $A_1+A_2 = 1$: the vertical normal stress is a conserved quantity
in $z$.) A ray of amplitude $A_i$, with velocity $c_i$, also carries
shear and horizontal normal stresses $\sxz = c_i\szz$ and
$\sxx = c_i^2\szz$. (Here $i=1,2$; these relations may be confirmed by
direct application of the
wave equations). Because shear stress is also conserved in
$z$ (for a point source of weight, the $x$-integral of $\sxz$ is
zero), one has $A_1/A_2 =
|c_2/c_1|$.

Since the wave velocities $c_1$ and $c_2$ become reversed as one crosses the
centreline, in all cases (except for the BCC model where $c_1=-c_2$), this line
forms a boundary between two different wave media and any ray impinging on it
undergoes both reflection and refraction. For simplicity we now move the
origin of our $z,x$ coordinates to the point where the ray meets the
centreline, in which case
an incident ray emanating from our point source corresponds to a disturbance
$\sigma_{zz} = A_2
\delta(x-c_2z)$, whereas the reflected ray obeys
$\sigma_{zz} = RA_2 \delta(x-c_1z)$, and  the transmitted ray
$\sigma_{zz} = TA_2 \delta(x+c_1z)$ (this incorporates the sign change of $c_1$
on crossing the centreline).  The factors $R$ and $T$ can be deduced as follows.
First, one imposes the conservation law for $\szz$ defined above;
the total weight supported by the reflected and transmitted rays is the same as
that in the incident ray. This yields immediately $R+T=1$. Secondly,
by considering the force on a small element, one finds that not only the shear
stress but also the horizontal normal stress $\sxx$ must be continuous
across the
centreline. (Note that the same does {\it not} apply, in general, to the
vertical
normal stress.)  Imposing this for the normal stresses, we equate
$\sxx (x=0^+) = A_2 c_2^2\delta(x-c_2z) + A_2 R c_1^2\delta(x-c_1z)$ 
with $\sxx(x=0^-) = A_2 T c_1^2\delta(x+c_1z)$. 
Using also the fact that $\delta(x-cz) = |c|^{-1}\delta(x/c - z)$, we find a second relation,
$|c_1/c_2| + R = T$. Thus we obtain the results
\begin{eqnarray}
T &=& \left(|c_2/c_1|+1\right)/2 \label{reftrans}\\ \nonumber
R &=&
\left(1-|c_2/c_1|\right)/2
\end{eqnarray} which completes our analysis of the
reflection/refraction processes. 
Note that for OSL models with dip ($\mu < 0$) the reflected ray factor 
$R$ has to be negative.

The above argument shows that the stress
response at height
$z$ to a point source above this level in the pile
consists of either two delta functions (amplitudes $A_1$ and $A_2$) or three
delta-functions (amplitudes $A_1, A_2R$ and $A_2T$) according to whether or
not the inward-going ray from the source has met the centre-line.
Using this information we can construct a geometrical solution of
the wave equations for each stress component; for
the vertical normal stress
$\sigma_{zz}(x)$ at a point $x$ on the base (say), this is done as
follows (Fig.5(b)). From each of two points separated by a small
distance $\Delta x$ centred on $x$, construct the backward light rays (allowing
for any reflection at $x=0$). This defines two strips of material, one of length
$L_1$ and the other of length $L_2$, with a third and fourth each of length
$L_3$ if there is a reflected/refracted ray. The corresponding widths
$w_1,w_2$ (and
$w_3=w_4$) are as shown in the figure. The total vertical normal force
between our
two points now obeys
\begin{equation}
\sigma_{zz}(x)\Delta x = g\left(A_1 (w_1 L_1 + R w_3L_3) +
A_2(w_2L_2+T w_3L_3)\right)
\label{rayweight}
\end{equation}
In other words, one adds the stress contributions of all the material elements
which have a light ray ending in the given interval
$\Delta x$ on the base. The above construction
provides a formula for $\sigma_{zz}$ which is, of course, identical to
eqns.(\ref{eqoutside},\ref{eqinside}) derived earlier.
Since the other stress components also obey a wave
equation, each of these can be constructed similarly, as a weighted sum of the
three $L$'s. The fact that each stress component is a piecewise
linear  function of $x$ then follows from the elementary geometry of triangles.

The Green function construction shown in Fig.5 gives some direct insight into
the role of the arching concept in describing stress propagation in
granular media.
The relation between the characteristic slopes $c_{1,2}$ and the
arching effect is far from intuitive, however. Specifically we can ask how,
starting from the BCC model ($c_1+c_2=0$) one should adjust the tilt
parameter ($\tau$) of the OSL model so as to obtain an
arching effect, and thereby a dip in the stress. In a slightly
different language\cite{language}, this was addressed briefly in
Ref.\cite{BCC}, where it was suggested that to get an arching
behaviour the light rays emanating from an element would have to be tilted
{\em outwards} ($\tau > 0$, or $c_1+c_2>0$), thus transporting the load away
from the centre. This suggestion, though at first sight
reasonable enough, is actually wrong. In the FPA model, and all other OSL models
giving a dip, the light rays are actually tilted {\em inward} relative to the
BCC model: their average slope,
$(c_1+c_2)/2$ is negative ($\tau < 0$). This would, at first sight, appear to
carry the weight of the grains {\em toward the centre of the pile}. The paradox
is resolved by realizing that, on tilting the rays inwards, the amplitudes
$A_1,A_2$ defined above, adjust so that $A_1$ becomes larger than
$A_2$. This means that a higher fraction of the weight of a grain is carried
along the outward ray, and away from the centre of the pile; this
redistribution is more than enough to compensate
for the average inward tilt of the two rays.

\subsection{Trollope's Model}
As mentioned in the Introduction, Trollope \cite{Trollope50s,Trollope} proposed
a model which yields, in effect, Edwards' arches and BCC as its two limiting
cases. The relation between this model and our own work is most clearly seen
in terms of the above analysis using rays; we therefore discuss it now.

In his model, Trollope, without invoking any differential formulation of the
problem, directly assumed that the stress could be constructed
in a manner similar to that above, but using {\it three} rays; two with equal
and opposite velocities $c_1=-c_2$ (just as in the BCC model), and a third,
horizontal ray
($c_3=-\infty$). The parameter $c_1$ was taken as
fixed globally by the type of packing; however, a
second parameter $k$ was introduced. This $k$ represents an imposed amplitude
ratio $A_2/A_1=k$ for the outward and inward components of the BCC-like
propagation; as $k$ is varied, the amplitude $A_3$ of the third ray also changes
(in a manner that can be deduced from stress continuity). It is
interesting that Trollope already realized the importance of singular behaviour
on the centreline; for $k\neq 1$, his model has this property.

In the limit $k=1$ (no arching, symmetric propagation), $A_3$ vanishes and one
recovers a BCC-like picture (though unlike BCC, Trollope did not
connect $c_1$ to the repose angle). In the limit $k=0$ one again has
two rays, one of which is now horizontal. This limit does not correspond to any
OSL model however: within the OSL model an infinite $c_2$ (representing the
horizontal ray) automatically has zero amplitude: $A_2=0$. Trollope's horizontal
ray enables stress continuity to be satisfied while giving a maximal
dip (zero  normal stress $\szz$ at $x=0$), reminiscent of the Edwards approach.
(The Coulomb yield criterion is violated, however.)

By use of the third ray, Trollope managed to interpolate these limits in what he
called the ``systematic arching theory". However, the introduction of this
extra ray seems extremely ad-hoc, which is perhaps why the model is not more
widely used today. Mathematically its presence means that rays can no longer be
identified as characteristics of a partial differential equation in two
dimensions; therefore Trollope's construction cannot
correspond to any local constitutive equation among stresses. (All such closures
must lead to hyperbolic equations for which a formal solution using
characteristics is available, even if the characteristics are curved -- as
happens, for example, in the IFE model \cite{nedderman}.)
We conclude that Trollope's systematic arching theory must be rejected as
unphysical -- a view tacitly shared by most of the recent sandpile literature.
However, many of the physical ideas behind the model, including the emphasis on
discontinuities in propagation across the centreline, remain highly pertinent
to the present work.

\section{The conical sandpile}

\label{sec3D}

We now extend our continuum modelling approach to the three dimensional
conical sandpile.

\subsection{One additional missing constitutive equation}

\label{secDmissing}
The conical pile is as shown in Fig.1(d); in addition
to the cylindrical coordinates $(z,r)$ introduced before, an azimuthal
coordinate $\chi$ is required.  Since we have
axial symmetry around the $z$-axis, the principal axes of the stress
tensor must include the azimuthal
($\chi\chi$) direction. (The orientation of this tensor can thus be fully
specified, as
before, by the inclination angle
$\Psi$ to the vertical of the major principal axis in the $r,z$ plane.)
Recalling that the stress tensor is symmetric, we therefore have
$\src=\scr=\szc=\scz=0$.  Hence the three dimensional conical pile has only
{\em one} additional independent stress component
\scc \ compared to the two dimensional case \cite{BCC}. The
stress continuity equation for a conical sandpile is
\begin{eqnarray}
\partial_r \srr + \partial_z \srz & = & \frac{\scc-\srr}{r} \label{eqpdeD3}\\
\nonumber
\partial_r \srz + \partial_z \szz & = & g  - \frac{\szr}{r} \\ \nonumber
\partial_{\chi} \sigma_{ij} & = & 0
\end{eqnarray}
The first two equations differ from those found earlier in two dimensions
by additional ``source terms", $({\scc-\srr}/{r})$ and $- {\szr}/{r}$
respectively, on the
right hand side.

Because of the high symmetry of the conical pile, closure of these
equations requires only that we find {\it two}
constitutive equations which together should determine any two of the
independent stress components in terms of the remaining
two.
Choosing the latter as before ($\srz$ and $\szz$) we refer to the resulting
equation for
$\srr$ as the primary, and that for $\scc$ as the secondary constitutive
equation. Note that symmetry
requires also
\begin{eqnarray}
\szr(r=0) & = & 0\label{eqcylindrical}\\ \nonumber
\srr(r=0) &= &\scc(r=0)
\end{eqnarray}

As our yield criterion for plastic failure of the granular material
we retain the Coulomb criterion \cite{Conical} for cohesionless
granular materials (which becomes a relation between principal stresses
in the $r,z$ plane). However, the Coulomb yield criterion is
essentially two-dimensional in character and gives no explicit information
on the circumferential stress
$\scc$. In common with previous authors \cite{footstress} we argue
nonetheless that this should vanish on
the free surface, as all the other stress components do:
\begin{equation}
\gcc(S=1)=0.
\label{eqby3}
\end{equation}
Note that if we were to use instead the Conical Yield criterion
\cite{nedderman,Wood,Conical} or a similar
(fully three dimensional) condition at the surface,
eqn.~(\ref{eqby3}) would not be an extra assumption.

The form of the new source terms in eqn.~(\ref{eqpdeD3}) is of interest.
If these remain relatively small everywhere, one can expect to find (independent
of the form of chosen for the secondary closure relation) qualitatively similar
results  to those obtained earlier in the two dimensional case. This scenario is
indeed fulfilled for the various different secondary closures tried below. In
any case, given that all stresses vanish at the surface (as just described),
these source terms become strictly negligible near the free surface of the pile,
which may therefore be viewed locally as having a planar two-dimensional
geometry. Accordingly, the IFS boundary condition is completely unaffected; the
stresses on the surface of a pile obeying IFS are still given by
eqns.(\ref{eqby2a},\ref{eqby2b},\ref{barmy}). It follows that (subject to
the usual scaling
assumptions, see below) eqns.~(\ref{eqabdonslope}) and (\ref{eqPsionslope})
still govern the asymptotic behaviour near the free surface. Thus the
relation between the repose angle $\phi$ and parameters in
the primary constitutive equation (such as the tilt angle
$\Psi$ in the FPA model, or the $\eta$ and $\mu$ parameters in OSL) remain as
they were in two dimensions.
\subsection{Scaling analysis}

\label{scalingassumpt}

As for the two dimensional case, by invoking the absence of an intrinsic
length scale we may demand that solutions of of the stress continuity
equation take the RSF scaling form, eqn.~(\ref{eqonlyH}). Substituting this into
eqn.~(\ref{eqpdeD3}) gives a set of ordinary differential
equations:
\begin{eqnarray}
\grr'/c + \grr + \gzr - \gcc - s \gzr' & = & 0 \label{eqDED3}\\ \nonumber
\gzr'/c + \gzr + \gzz - s \gzz' & = & 1
\end{eqnarray}
The asymptotic analysis in Section 2.4 for the stresses
near the surface carries over to the case of three dimensions, as mentioned
already above.
The source terms in eqn.~(\ref{eqpdeD3}) can in principle affect the the
asymptotic behaviour given in Section 2.4 near the centre of the pile (i.e.,
small
$S$) but {\it qualitative} changes arise only if either $\szr/r$ or
$(\srr-\scc)/r$ become large in this limit. This does not occur for any of the
models studied below.

Following the arguments made earlier in two dimensions, based on our
assumptions of RSF scaling and ``perfect
memory'', we now propose local
forms for both the primary and secondary constitutive relations, which must
be as
follows:
\begin{eqnarray}
\grr/\gzz & = & \Closure(U) \label{eqClosureD3}\\ \nonumber
\gcc/\gzz & = & \ClosCon(U)
\end{eqnarray}
where we have set $U=\gzr/\gzz$ as usual.

The form of eqn.~(\ref{eqRK}) for the RSF scaling solution remains
basically unchanged, except that in eqn.~(\ref{eqabcdef}) the terms
$\ccc=\gzr+\grr-\gcc$ and $\fff =
\gzz +\grz -1$ are somewhat modified.
The resulting stress profiles can readily be calculated
numerically from eqns.~(\ref{eqRK},\ref{eqabcdef}) for any choice of the
closure relations.
However, no analytic solution appears to be obtainable even for those models
which, in two dimensions, reduce to wavelike propagation. The problem
can, of course, still be viewed as quasi two-dimesional (one spacelike, one
timelike variable), but if so the extra ``source terms" make the solution
complicated. Alternatively these models can be formulated in terms of wave
propagation in two spacelike dimensions (the $r,\chi$ plane) with $z$ as a
timelike variable. However, the Green function for such waves is itself
surprisingly complicated (there is no sharp ``light-cone"  \cite{BCC}) and not
directly amenable to the simple geometrical interpretations offered earlier.

\subsection{Choice of constitutive equations}

For the primary constitutive equation, we can choose among those
discussed earlier, namely IFE and OSL, with the latter including both BCC
and FPA
as special cases. We continue to require that the IFS boundary condition is
obeyed at the surface (which again fixes the OSL parameters $\mu$ and
$\eta$ to lie on the IFS line) and that the Coulomb yield criterion
is not violated in the interior of the pile.

For the secondary constitutive equation, we have investigated three ways of
selecting the function $D(U)$. The first is to insist that $D(U)$ coincides with
$C(U)$ so that $\scc = \srr$ everywhere in the pile:
\begin{equation}
\Conicalone(U) \equiv \Closure(U)
\label{Closure1}
\end{equation}
This has the merit of simplicity. (Note that by symmetry this relation must hold
anyway at the central axis of the pile, but not necessarily elsewhere.)  A
second choice is suggested by the observation that the $\chi\chi$ direction is a
minor principle axis for a conical sandpile with axial symmetry. Generalizing
slightly an assumption often made the context of conical
hoppers \cite{Wood,nedderman}, one could then choose as the secondary closure
$\scc=P-R$ (the Haar - von Karman hypothesis \cite{Wood}). This implies
\begin{equation}
\Conicaltwo(U)=(1+\Closure(U))/2-\sqrt{(1-\Closure(U))^{2}/4+U^2}
\label{Closure2}
\end{equation}
Although the motivation for this choice in the sandpile context is not very
clear, we have tried it out for comparison.
Our third choice of secondary closure, unlike the first two, does not
explicitly depend involve the primary closure $C(U)$; it is the linear
relation
\begin{equation}
\Conicalthree(U)=\eta + \tilde{\mu} U
\label{Closure3}
\end{equation}
which should be compared with the OSL primary closure,
eqn.(\ref{eqlinclosure}). In fact,
for the OSL model the constant term $\eta$ has to be identical to that
chosen in the primary closure, to meet the second
requirement of eqn.~(\ref{eqcylindrical}). The coefficient
$\tilde{\mu}$ is in principle free. In practice, however, we have found that
the requirement that the Coulomb criterion $\Yield \leq 1$ holds in the
interior of the pile means that values of
$\tilde{\mu}$ close to $\mu$ are required; hence for OSL models the
closure $D_3(U)$ is never very different from $D_1(U)$.

We have investigated these three closure relations $D_1,D_2,D_3$ for all the
different primary closures \Closure \ already discussed in Section \ref{sec2D},
for various values of the repose angle $\phi$ (mainly in a range around $\phi =
30^o$). For all the parameters we tried, the extra ``source terms"  led mainly
to smoothing of the two dimensional curves without qualitatively altering the
presence or absence of the dip. Since these source terms do not have a dramatic
effect, it follows that the choice made for $D$, at least among those
investigated here, itself does not qualitatively change the stress profiles.

\subsection{Results}

Rather than provide a catalogue of curves for various combinations of
primary and
secondary closure, we will focus attention on the FPA model.
In Fig.6 we compare the stress curves for the
three-dimensional FPA with closures $D_1$ and $D_2$. As mentioned
previously, the choice of secondary closure proves quantitatively but not
qualitatively important.
Also shown are the
experimental results of Ref.~\cite{smid} for piles of height 20-60 cm.
The stresses are normalized by the total weight of the pile; notice the good
scaling collapse of curves from piles of varying heights. This confirms that the
RSF scaling hypothesis made in this paper is obeyed to experimental accuracy, at
least for the materials and pile sizes studied in Ref.~\cite{smid}.
The agreement between experiment and FPA theory is generally
satisfactory, although there is a significant error near the maximum of the
vertical normal stress. Obviously it would be helpful to have more data for
small values of $S$, but there are sufficient data points at the origin to
clearly establish the presence and magnitude of the dip.
The experimental data
shown are for two different media both with repose angles close to $\phi =
33^o$. The resulting curves differ by an amount similar to the difference
between
the two choices of secondary closure, with $D_1$ giving slightly better results
for  ``quartz sand" and $D_2$ for ``NPK-1 fertilizer". (We do not attach any
significance to this.) As mentioned previously, the FPA model has no adjustable
parameters once
$D$ is chosen and $\phi$ is fixed by experiment.

In Fig.7 we show the same predictions for the FPA model with closure $D_1$
alongside those for several other models with the same secondary closure. These
models are BCC and IFE (neither showing a dip); and two parameter choices for
the OSL model ($\eta = 0.8$ and $\eta = 1.2$) which bracket the FPA case
($\eta =1$). This comparison shows a
clear preference of FPA over those other models that have no adjustable
parameter. It is conceivable that the data could be fit better by choosing an
OSL model with $\eta$ slightly different from unity. However, we do not believe
the improvement is enough to justify the adoption of an extra fitting
parameter, although further careful experiments might reveal this to be
necessary.

Finally in Fig.8 we plot the yield parameter
$\Upsilon(S)$ for the BCC, FPA and IFE models.
By definition,
$\Upsilon(S) = 1$ everywhere in the IFE model; it also obeys $\Upsilon(1)
=1$ in all
models obeying the IFS boundary condition. In two dimensions it is also unity
throughout the outer regime of the pile for all OSL models. In three
dimensions this is not the case, and in fact for OSL models the material is
clearly below
the yield criterion throughout the bulk of the pile. This underlies the
important distinction between the classical IFE assumption (fully mobilized
friction, $\Upsilon = 1$) and the new models adopted in this paper.

\section{Role of construction history}
\label{secalpha}
So far we have only considered the stress profiles of idealized sandpiles
constructed by pouring sand from a stationary pipe, for which the boundary
condition of incipient failure at the surface (IFS) was
assumed.
The slope $\alpha$ to the horizontal of the free surface is
by definition given by the repose angle: $\alpha=\phi$.
According to our approach, however, the constitutive equation encodes the
construction history, and piles built differently can behave differently.
In discussing this issue, we restrict attention to the FPA constitutive model in
two dimensions.

We consider first the following hypothetical experiment: a
material with $\phi = \phi_0$ is formed into a pile by the usual method. The
pile is then reduced to a flatter (symmetrical) one of angle $\alpha$ by simply
taking away the upper section, grain by grain, without disturbing any material
below (Fig.9(a)). According to our model, the constitutive equation remains that
of a pile with the larger repose angle, though the stresses are of course
altered. The resulting stress pattern is shown in Fig.9(b) in comparison to that
of a pile of repose angle $\phi_1=\alpha$ which has the same final geometry. The
first of the two piles has the larger dip.
We can now ask the following: if a pile of $\alpha < \phi$ is
tilted from the base through an angle $t$ (Fig.9(c)) how large may $t$ become
before an avalanche occurs? A classical answer, based on the view that
the repose angle $\phi$ is a material property {\it independent of construction
history}, is that one would
be able to tilt until $t_{max}+\alpha$ is again equal to $\phi$.
(This ignores, as we have done throughout this paper, the small hysteresis
effects
associated with the Bagnold angle \cite{Bagnold}). However, in our approach this
should not quite be true, since the inclination angle of the principal axes in a
pile at this condition is different (by an angle $t$) from that of a pile
created by the normal method at its repose angle.
It turns out, however (Fig.9(d))
that unless the pile is substantially flattened ($\phi-\alpha \simeq 10^o$ or
more), the difference between $t_{max}+\alpha$ and $\phi$ is very slight, at
least for repose angles in the usual range ($\phi < 45^o$).

This calculation can, with caution,
be proposed as a model for what happens when a sandpile, built normally,
is suddenly tilted through a finite angle $t$. Of course, in this case an
avalanche does occur: however, if this happens by removal of a wedge {\it
without significant reorganization of the remaining part of the pile}, leaving
the new surface in a state of incipient failure,
the above calculation can be applied (except that, for simplicity,
we have contrived a version in which the pile remains symmetrical). In
principal, there should then be a change
in the resulting repose angle if $t$ is large enough. However, the assumption
that an avalanche occurs with no rearrangement of the remaining grains, is, for
large $t$, highly dubious.

Another critical test of our ideas is the following: a
triangular pile is constructed as usual and then a large part of it removed
(grain by grain) leaving a pile whose left hand slope is at the angle of repose
$\phi$, and whose right hand slope is at angle $\beta$ (say) to the horizontal.
The geometry is chosen so that all of the material in the new pile was
{\it originally in the left half} of the parent pile (Fig.10(a)).
To describe this situation, we have to use the FPA constitutive equation
in the form (\ref{fixedpaxz}), in a coordinate system where $x=0$ denotes
the centreline of the original pile.
With the modified construction history just described, the
singularity on the line $x=0$ lies outside the newly created pile, and the
characteristics of stress propagation should be identical on both sides of its
apex. Throughout the pile, a majority
of the stress is carried down the {\it leftmost} (rather than {\it outermost})
characteristic. An interesting question now is,
what is the maximum angle $\beta$
that we can choose for the right hand slope?
This can be found from the usual
stability criterion
$\Yield\le 1$; the marginal case has equality at the free surface on the right
and, for the FPA model, this gives (after some algebra) the condition
\begin{equation}
\tan(\psi - \beta) \ge \tan^3(\psi)
\label{tangent}
\end{equation}
The maximum $\beta$ for which the new pile is stable is shown
as a function of $\phi$ in Fig.10(b); for $\phi = 30^o$, one has $\beta_{max}
= 19.1^o$. This is a very interesting result, since it predicts that the repose
angle of the right hand part of a pile built this way is quite different from
the usual value, which prevails on the left.

This prediction must, of course, be
interpreted with caution since its extension to a fully three dimensional
geometry is not obvious. Perhaps the simplest three dimensional analogue is to
build a pile and then open a hole directly below the vertex, allowing sand to
flow out leaving a ``volcano crater" \cite{entov}; according to this prediction,
the angle of repose on the inner side of the crater may differ substantially
from that on the outer slope. This possibility deserves careful
experimental study; a significant difference is not ruled out
\cite{Rotter}. The situation is
again complicated by the fact that the experiment will set up a flow which may
rearrange the grains that remain in the pile. Indeed,  the
removal by avalanche of the right hand part of the pile may set up a large
region in which the grains have slipped down to the right, for which
the constitutive equation may revert to that of the right hand part
of a normal pile. (This could be true even if the actual particle displacements
are extremely small.) If so, the measured repose
angle could again approach $\phi$, rather than $\beta_{max}$ which applies
only when the removal of sand does not perturb the remainder.
We show in Fig.10(c) the stress distribution in a pile
made in this careful fashion (with $\phi = 30^o$ and $\beta = 15^o$). As one
might expect, there is now  no dip but a (lopsided)
maximum in the vertical normal stress. The maximum lies to the left of the
new apex (at the point where an outgoing ray from this apex strikes the base).

Note that quite different predictions for this geometry could have been
obtained by writing the FPA closure in a somewhat different
form, which is, {\it for a symmetrical pile only}
equivalent to eqn.~(\ref{fixedpaxz}) \cite{nature}, as discussed in Section
2.7:
\begin{equation}
s_{xx}/s_{zz}
= 1 - 2\tan\phi\, |U|
\end{equation}
Here the explicit dependence on construction history
via the $\sign(x)$ factor has been replaced by an $x$-independent but
highly nonlinear
constitutive relation among the stresses (in the spirit of some of the models
discussed in Ref.\cite{BCC}). Using this form, one could find a solution,
with $\beta = \phi$, for our asymmetrically constructed pile
that would precisely coincide with the usual symmetrical case. (This possibility
arises because the sign of $U$, though not of $x$, can change on the centreline
of the new pile.) Since we only have data for symmetric piles, we cannot
on the existing facts rule out this rather different version
of the FPA model, although it does not correspond to our assumption that
the principal axes
of a material element are fixed at the time of burial. Accordingly
experiments on asymmetric piles would be a strong test of the theory.

A somewhat different experiment would be to start with a symmetrical pile and
then remove parts of it (grain by grain) so that the remainder forms an
asymmetrical pile whose apex has not moved from the line $x=0$. Shown in
Fig.11 is the stress distribution in such a pile with left and right
slopes $\alpha_1 = \phi = 30^o$ and $\alpha_2 = 12^o$. An interesting
feature is visible under the apex, where the vertical normal stress
$\sigma_{zz}$ (and therefore also the yield function $\Yield$) is discontinuous.
This behaviour is in fact a generic feature of sandpiles that include
the line $x=0$ but are asymmetric about it. It stems from the fact that
this normal stress is not continuous in the geometry of incident, reflected
and transmitted rays considered earlier (Section 2.9).
An exception to this rule is if an asymmetrical pile is made by pouring sand
onto a sloping base plate. In this case,
we expect the apex of the pile to move slightly relative to the source so that
more material rolls down the ``long" side of the pile and the repose angle in
the two directions remain equal to $\phi$. The constitutive equation must
then (given RSF scaling) be the same as for a pile formed normally, and the
stress exerted on
the supporting plate is the same as that on an inclined plane inscribed
through a
normal pile. (In two dimensions, this can be found easily from our earlier
results.) Though asymmetric, this stress distribution will not show any
discontinuity beneath the apex.

As a last example of a sandpile constructed normally and then
manipulated, in Fig.12 we show the stress distribution in a pile whose top
section
has simply been cut off. Though grain-by-grain removal of sections of a pile may
impose experimental difficulties, this is perhaps the simplest geometry for
which it could be achieved. As shown, the dip is gradually diminished and
replaced by a plateau as larger and larger upper sections are taken away.

Finally, we note that a sandpile could be made by first distributing sand
uniformly (not from a point source) in a retaining bin, from which the
side walls are then removed. As with some of the examples studied above, the
predictions depend crucially on whether significant slip occurs within
the part of the pile that finally remains. The initial loading
of the bin is likely to produce principal axes with vertical and horizontal
orientations ($\Psi = 0$), so that if no slip occurs, we would
expect the BCC model to apply (no dip). However, if slip does occur so the
remaining pile has been sheared downwards, the FPA picture should be more
appropriate.

The various types of experiment discussed above, in which the construction
history of the pile is deliberately manipulated, provide a strong test
of our basic modelling hypothesis that the constitutive equation encodes the
construction history. For some of these geometries, the theoretical
predictions challenge the ``classical" assumption, maintained in the recent
physics literature on sandpiles,
that for cohesionless granular media (of a single grain size \cite{nedderman})
the repose angle is the same for all types of pile of a given material.
(As mentioned in Sec. 1.2, this assumption has long been avoided
in the engineering literature on hoppers \cite{Rotter,rotter1}.)
Of course, the repose angle
remains a genuine material parameter in that the angle of a ``normal" pile
(built from a point source) will differ for different cohesionless materials;
and for our purposes this can be taken as the unique definition of $\phi$. It
does not necessarily follow, however, that the repose angle taken up by a pile
of the same material with a different construction history, will always be
exactly the same. In any case, our modelling approach leads to a clear
expectation that the {\it stresses} in such piles can be different (even
when the
repose angles are not significantly different). This is a readily testable
prediction which we believe deserves urgent experimental attention.

\section{Conclusion}

This paper is a long one. It therefore seems useful to provide a brief summary
of our modelling strategy, in the form of a list of
contentions for which more detailed arguments can be found in the text
above. We stress, however, that several points on the list have no
first-principles justification: they are hypotheses whose value can at
present only be judged by comparison with experiment. We also stress that
several of these ideas have a long history (which is not the same as
saying that they are widely agreed upon).

\subsection{A manifesto for sandpile modelling}
Our modelling strategy is based on the following claims:

(1) There is a construction history, $\cal H$. This determines the
arrangement of grains. We define the ``normal" history to be the construction
from a point source of a pile at its repose angle.

(2) There is a stress tensor $\sigma_{ij}$ which is well-defined as a
local (mesoscopic) average over many grains.

(3) For hard particles (of infinite elastic modulus), no strain variables
exist; static frictional forces are indeterminate.  Stress continuity
requires one supplementary equation for closure in two dimensions, and two for
a conical pile in
three dimensions.

(4) Scaling behaviour (RSF scaling) is observed, to experimental accuracy.
Hence there is no characteristic length scale in a sandpile under
gravity. Particle deformability would provide such a length; so would size
segregation.

(5) The limit of uniform nondeformable, cohesionless, particles presumably
therefore
exists, and should describe those experiments for which RSF scaling is observed.

(6) We should therefore seek as closure a scale-free, local constitutive
relation
among stresses.
Formally: there is
a function  $\cal C$ such that
\begin{equation}
{\cal C}(\sigma_{ij}(r,z), {\cal H}) = 0
\end{equation}
The constitutive relation depends on the local packing and therefore
on the construction history: $\cal C$
encodes $\cal H$.

(7) $\cal C$ for a material element is ``frozen in" at the time of burial
(perfect
memory assumption). Combined with RSF scaling, this means that for a sandpile
constructed from a point source, $\cal C$ is independent of
position when expressed in cylindrical polar coordinates,
though it may be singular on the
central axis.

(8) The boundary conditions for a pile constructed normally are IFS: incipient
failure at the free surface. This means that {\it at the surface}, the major
principal stress axis bisects the free surface and the downward vertical. (Here
and elsewhere,
hysteresis effects associated with the Bagnold angle are ignored.)

(9) The search for a constitutive relation ${\cal C}(\sigma_{ij}, {\cal H})$
may legitimately entail (a)  making
simplified hypotheses to compare with experiment;
(b) microphysical modelling from first principles. We pursue the former
in this paper, the latter elsewhere \cite{volkard2}.

(10) A classical choice of $\cal C$ is incipient failure everywhere (IFE);
this is hard to defend physically. It does not predict a dip in the stress
beneath the apex of a pile.

(11) A physically more plausible (but by no means unique) choice for $\cal C$ is
provided by the FPA hypothesis. According to this, each element of
material is impressed at burial with a sense of direction, which fixes forever
the orientation of the stress tensor ellipsoid that the element can support.
The model predicts a dip in two dimensions.

(12) In three dimensions, a secondary closure relation is needed. Among the
more obvious choices, it makes relatively little difference which
is chosen. Even the simplest choice ($\scc = \srr$), combined with FPA, gives
a reasonably good fit to the data of Ref.\cite{smid}, without adjustable
parameters.

(13) A generalized model (OSL), of which FPA is a special case, can be
introduced. This has an adjustable parameter, the introduction of which is not
demanded by the present data.

(14) The above modelling approach, though initially set up for static sandpiles
constructed from a point source, can also be used for more complex construction
histories (at least in some cases). For piles constructed normally and then
modified by careful removal of grains, this approach predicts a nontrivial
dependence of the repose angle $\phi$, and of the stress distribution, on
the way a pile is made.

\subsection{Discussion}
Of the models considered in this paper, it is clear that the FPA model
has some especially attractive features.
This model leads directly to an arch-like
stress-propagation, with the major part of any load being carried down the arch
direction. The latter coincides with the major principal axis of the stress
tensor; this {\it everywhere} bisects the free surface and the downward
vertical. The predictions of the FPA constitutive relation thereby describe
similar physics to the arching model of Edwards \cite{EO} (and indeed the
earlier ``full arching theory" of Trollope \cite{Trollope}). Like such models,
the FPA hypothesis can be viewed as a direct macroscopic ansatz of how stresses
propagate: one assumes that the principal axes are fixed in space.
Viewed this way, we believe that the FPA model provides the first
description of the arching picture within a fully consistent continuum
mechanics framework.
Its experimental success strongly
suggests that the presence of a macroscopic arching structure in sandpiles is
the correct explanation for the observed minimum in the vertical normal stress
below the apex of the pile.

However, unlike previous arching models, the FPA hypothesis can also
be interpreted as providing a local, history-dependent constitutive
relation among
stresses. In this context, it is among the simplest such equations that can
plausibly be devised: we assume that the principal axes of a material element
are fixed at the time of its burial. Viewed as such,
the FPA hypothesis contains no assumption of any macroscopic arching structure;
rather, it provides a plausible microscopic explanation for how such structures
arise. Its experimental success offers strong support for a modelling strategy
cast in terms of such constitutive relations.
For parameter values other than FPA, which is a special case, the more
general OSL model predicts, within the same modelling framework, a more
complex pattern of stress propagation. (The principal axes and the
propagation characteristics no longer
coincide.) The extra fitting parameter provided by OSL is probably
not justified by the existing data. One feature of OSL models which
stands out strongly (at least in two dimensions) is the presence of
reflection and refraction of stress-paths at the central axis of the pile
(see Sec. 2.9). Careful experiments on the effect of small perturbing
loads could reveal whether or not this really occurs, providing a strong
test of this class of model.

In view of its attractive physical features, and of its experimental success,
we currently favour the FPA
hypothesis as the simplest starting point for more refined theories of
sandpiles. It also forms a promising basis for future study of stress
propagation
in static granular media of geometries quite different from the normal conical
pile. The richness of this area is amply illustrated by the handful of examples
studied above in Section 4. Consideration of these and other geometries
could allow
stringent experimental tests of both
the FPA model, and the overall modelling strategy
we have proposed. Within this framework,
there is, no doubt, scope for much more sophisticated models of how the
construction history of a pile determines the local constitutive behaviour,
but further efforts in this direction may require much more
experimental input. The validity of the framework itself deserves close
experimental scrutiny, particularly concerning the degree to which RSF scaling
is obeyed. Our assumption of scale-free (RSF) behaviour offers an immense
simplification, but closer experimental investigation may reveal that this
is not
quantitative except under some limiting conditions.
Despite these uncertainties, we feel that the modelling framework presented
above has significant potential to provide improved physical theories of stress
propagation granular media.

In future work \cite{ournoise} we will explore the close
connection between our OSL model and a
recent discrete stochastic models for stress propagation in sandpiles
\cite{degennes} (see also \cite{Liu}), of
which OSL can be viewed as the (mean-field)
continuum limit. An important concept arising from the stochastic models
and from experiment \cite{Liu} (see also \cite{nagel}) is that of {\it
stress paths}; these are pathways through the medium along which most of
the load
is locally transmitted. The noise-free models considered in this paper
can be viewed as making hypothetical statements about the {\it average}
orientation and load-bearing properties of these paths (see the discussion
of characteristics in Sec. 2.9). Such statements are testable, if not directly,
then at least in simulation studies. Ongoing work \cite{volkard2} suggests
a promising correspondence between the average alignment of these paths
and the orientation angle $\Psi$ arising in the OSL model.

$\,$

{\bf Acknowledgement:}
The authors are indebted to J.-P.~Bouchaud for a series of essential
discussions which
laid the foundations for this work. We also thank S. F. Edwards, V. Entov,
M. Evans, J.
Goddard, T.C.B. McLeish, R. M. Nedderman and J. M. Rotter for illuminating
discussions. JPW
thanks C.~E.~Lester, C.~S. and M.~J.~Cowperthwaite for supporting
enthusiastically this research. MEC acknowledges the hospitality of the
Isaac Newton Institute
for Mathematical Sciences (Cambridge) where part of this work was done.
This work was funded in part by EPSRC under
Grant GR/K56223 and in part by the Colloid Technology programme.

\newpage

\begin{figure}
\caption[]{The symmetrical sandpile.
(a) Definition of the normal construction history of a pile.
The grains fall down from the point source on the pile and
roll down the slopes, which are at the repose angle $\phi$.
(b) The arching concept.
In the Edwards-Oakeshott formulation the weight supported at
a point the base is proportional to the length of the arch
impinging on that point. Outer (incomplete) arches are unstable.
(c) Coordinates for the 2-d sandpile. The scaling variable $S=r/(cz)$
is unity on the free surface. The height of this pile is $H$.
The $(z,r)$ coordinates, and also a second set $(n,m)$ rotated
through angle $\tau$ are shown. The ellipse denotes the
stress tensor whose major axis is inclined at angle $\Psi$ to
the vertical in the neighbourhood shown.
(d) Cylindrical polar coordinates for the 3-d pile.
}\end{figure}

\begin{figure}
\caption[]{
Results for a two dimensional symmetrical sandpile with a surface obeying IFS;
$\phi=30^0$. (a) Reduced shear stress $s_{zr}$ and reduced vertical
normal stress $s_{zz}$ as a function of scaling variable $S=r/(cz)$.
Results for the IFE model (found numerically), the BCC model, and the
FPA model are compared. Those for the third stress component $s_{rr}$
are not shown but can be deduced from those given via the appropriate
constitutive equation in each case.
(b) The same comparison, showing instead the yield function $\Yield(S)$.
For IFE this is unity everywhere by definition; for FPA and BCC in two
dimensions it is unity throughout the ``outer" regime of the pile (the
same does not apply in 3-d). In the FPA model the stress at the centreline
is isotropic and $\Yield =0$ there. (c) The same comparison, showing now
the orientation angle $\Psi$ of the major principal axis. At the free surface,
where $\Yield = 1$, $\Psi$ bisects
the free surface and the vertical: $\Psi = \psi$ (which is $30^o$ in this
case). The same relation holds everywhere in FPA; in BCC it holds only
in the outer regime. In the IFE model, it holds at the surface only.
The ``shooting'' to the surface value in the numerical solution of
IFE is not perfect because of the numerical instability
generated by the singularity on
the surface.
}
\end{figure}

\begin{figure}
\caption[]{Sketch of the geometry of the FPA model. The stress ellipsoid
has fixed inclination angle $\Psi = \psi$; its ellipticity varies from zero
at the
centre of the pile to a maximum in the outer region. The outward and inward
stress propagation
characteristics are indicated by short-dashed and long-dashed lines; these
are at rightangles and coincident with the principal axes of the stress
ellipsoid.
}
\end{figure}

\begin{figure}
\caption[]{The $(\mu,\eta)$-plane for OSL model parameters. For a normal
pile these must lie on the IFS line, shown as a full line for $\phi = 30^o$
and dashed for $\phi = 10^o$.
The BCC model (open symbols) and FPA model (filled symbols)
are marked on the IFS line in each case.
Note that, on this plot, BCC models (for different $\phi$) all lie on the
vertical
$\mu=0$ axis, separating models showing a dip ($\mu<0$) from those with
a hump. Likewise FPA models (for different $\phi$) all lie on the
horizontal $\eta=1$ axis, separating models active near the centre of
a pile (below the axis) from those which are passive there ($\eta >1$).
Dash-dotted lines denote the zone within
which the solutions obey the Coulomb yield requirement; the left boundary
(marked $R_P$) denotes passive failure at the centre of the pile and the
right ($R_A$) active failure there.
}
\end{figure}

\begin{figure}
\caption[]{Construction of the solution of the OSL model using characteristics.
(a) The response to a point source is constructed as three rays consisting
of delta functions of the amplitude shown. Reflection and refraction of the
rays at the centreline occurs when the two wave speeds, $c_1$ and $c_2$, are
not equal in magnitude.
(b) The solution for a symmetric pile is constructed by summing over all sources
whose rays end in a short segment $\Delta x$ at the base of the pile. This
defines
four strips of material, as shown, of lengths $L_{1-4}$ and widths $w_{1-4}$.
Multiplying the area of each strip by the appropriate amplitude factor,
and adding, gives the piecewise linear solution for the vertical normal stress.
Similar solutions for the other stresses are likewise obtained (using the
ratios of stress components within each ray as given in the text).
}
\end{figure}

\begin{figure}
\caption[]{Results in three dimensions for the FPA model for $\phi = 33^o$
with secondary
closures $D_1$ and $D_2$ defined in the text. Also shown are the data of
Ref.\cite{smid} for quartz sand (closed symbols) and NPK-1 fertilizer (open
symbols), both
of which have repose angles of $33\pm 1^o$. Though departures are apparent
near the maximum of the vertical stress, the dip is reproduced satisfactorily;
there are no adjustable parameters in the FPA model. The difference between
the two secondary closures is similar to that between the two materials,
though we attach no special significance to this. Note the data collapse from
piles of different heights; this confirms that RSF scaling is obeyed to
experimental accuracy.
}
\end{figure}

\begin{figure}
\caption[]{Comparison of different primary closures with the same secondary
closure relation $D_1$, for $\phi = 33^o$. The IFE and BCC models, which do
not give a dip, are clearly ruled out by the data of Fig.6. However, it is
harder to distinguish OSL models with the adjustable $\eta$ parameter in the
range $0.8 < \eta < 1.2$ from the FPA model which has $\eta = 1$. A parameter
values different from 1 cannot be ruled out, but nor does one seem to be
supported by the data of Fig. 6.
}
\end{figure}

\begin{figure}
\caption[]{The yield function $\Yield$ as a function of scaling variable
$S$ for the FPA, IFE and BCC models. Note that for FPA and BCC,
$\Yield$ now saturates
the Coulomb condition, $\Yield = 1$, only at the free surface and not
through a finite part of the pile. The apparent cusps on the FPA and BCC
curves are numerical artefacts arising from the shooting procedure used
to solve the equations.
}
\end{figure}

\begin{figure}
\caption[]{A pile whose shape is changed after being constructed normally.
(a) Geometry of the altered pile; dashed lines show the major principal
axis orientation, which is unaffected by the removal of grains above. (b)
Resulting stress distribution
compared with a pile whose repose angle is the same as the final one of
the altered pile. (Initial slope $\phi = 30^o$,
final slope $20^o$. ) (c) The application of a tilt $t$; (d) the final ``repose"
angle $\alpha + t_{max}$ of the tilted pile, plotted against $\phi$.  This
is determined by finding the
maximum $\alpha$, given $t$, for which the pile remains within the Coulomb
yield threshold. For reasonable
$\phi$, and small $t$, the final repose angle is almost the
same as that of a normal pile.
}
\end{figure}

\begin{figure}
\caption[]{Construction of an asymmetric pile from one half of a normal pile.
(a) Geometry of the altered pile; dashed lines show the major principal
axis orientation, which is unaffected by the removal of grains above. The
newly created pile has axes uniform throughout, rather than discontinuous
at the centre line. This alters the stress propagation behaviour. (b) The
maximum $\beta$ that can be chosen in the geometry of (a), as a function
of $\phi$, to avoid violation of Coulomb's yield condition in the newly formed
pile. According to the FPA model, $\beta_{max}$ is quite different from
the ordinary repose angle $\phi$. (c) Resulting stress distribution and
yield function
under the new pile (the scaling variables are the same as for the
unmodified pile).
The apex of the new pile is marked with an arrow.  There is now a lopsided
hump, rather than a dip, in the vertical normal stress.
}
\end{figure}

\begin{figure}
\caption[]{Construction of an asymmetric pile from both halves of a normal pile,
leaving the apex in the same position. (a) Geometry of the altered pile;
dashed lines show the major principal
axis orientation, which is unaffected by the removal of grains above.
(b) Resulting stress distribution and yield function
under the new pile (the scaling variables are the same as for the
unmodified pile).
The apex of the new pile is marked by singularities in the vertical normal
stress and in the yield function, for the reasons discussed in the text.
For the parameters shown, the dip is present in one half of the pile but
not the other.
}
\end{figure}

\begin{figure}
\caption[]{Vertical normal stress beneath a symmetric
pile made normally, of height $H$, from which
an upper pile of height $z_c$ has been removed. The dip is progressively
eliminated as material is taken away from the upper part of the pile.
}
\end{figure}
\vfil\eject


\begin{thebibliography}{99}
\bibitem{language}
In Ref.\cite{BCC}, a family of nonlinear models are produced in which
a velocity-like parameter $V$ is considered to be a function of the
shear stress. The OSL model can be viewed as the
limit in which $V(\sigma_{zx})$ approaches a step function (giving a
positive or negative constant value according to the sign of $x$).
A finite slope of this function at the origin ($V'(0)\neq 0$) would
lead to rounding of the cusp at the centre of the dip on some
characteristic length scale (in violation of RSF scaling).
\bibitem{deflection} If support deflection were important, one would
not expect RSF scaling as defined below. (The elasticity of the support
introduces a length scale just as that of particles would do.)
\bibitem{footdistrib}
In principle one could construct piles in which
the scaling was nonetheless obeyed, by demanding that the {\em relative}
variation of the material properties between (say) an element at the
apex, and an element directly beneath the apex at the base, must be the
same for all piles. However, this would require the spatial gradient of the size
distribution, for the element at the base, to vary inversely with the height
of the pile.  Clearly, the size distribution (and its gradient) in this
element is fixed forever long before the eventual height of the pile is decided.
We conclude that, although size segregation may arise, models invoking this to
explain the dip are inconsistent with the experiments of Ref. \cite{smid}.
\bibitem{discfoot} It turns out that
discontinuous derivatives can arise for three values of $S$: at the outer slope
($S= 1$); on the central axis of the pile ($S=0$); and the point $S=S_0$ where
the stress information from the apex of the pile reaches the bottom; examples
can be seen in Fig.2(a).
\bibitem{radialstressfield}
Obviously one can equally write the scaling ansatz
eqn.~(\ref{eqonlyH}) as $\sigma = g R Q(\theta)$ where
$R$ is the radius measured
from the top of the pile and $\theta$ is the angle measured
clockwise from the (downwards pointing) $z$-axis
(see Fig.1(c)).
This yields the trivial identification $Q(\theta) = g(S)/\sqrt{1+(cS)^2}$.
\bibitem{overdetermined}
It might seem that the problem is overdetermined,
but two equations
(the first of the eqns.~(\ref{eqabdonslope}) and eqn.~(\ref{eqby2a})) are
automatically identical in a radial stress field.
\bibitem{tilt} In the FPA model the tilted axes must be chosen so as to
coincide with the principal axes themselves.
\bibitem{midslip}
Though an analytic proof should be possible,
we have so far only checked this numerically.
\bibitem{Conical}
The ``Coulomb yield criterion'' is the frictional analogue of
Tresca's criterion \cite{Wood}.
The frictional analogue of von Mises' criterion is the ``Conical yield
criterion''. For a two dimensional problem both criteria are identical.
In three dimensions, the failure conditions predicted by these two
criteria only differ slightly and the stresses
(calculated for instance with an IFE assumption)
are predicted to be similar \cite{nedderman}.
\bibitem{footstress} One argument for this is that grains at the surface cannot
be pressed together more in $\chi$ direction than in $r$-direction:
$\grr(1) \geq
\gcc(1)$.
\end{thebibliography}
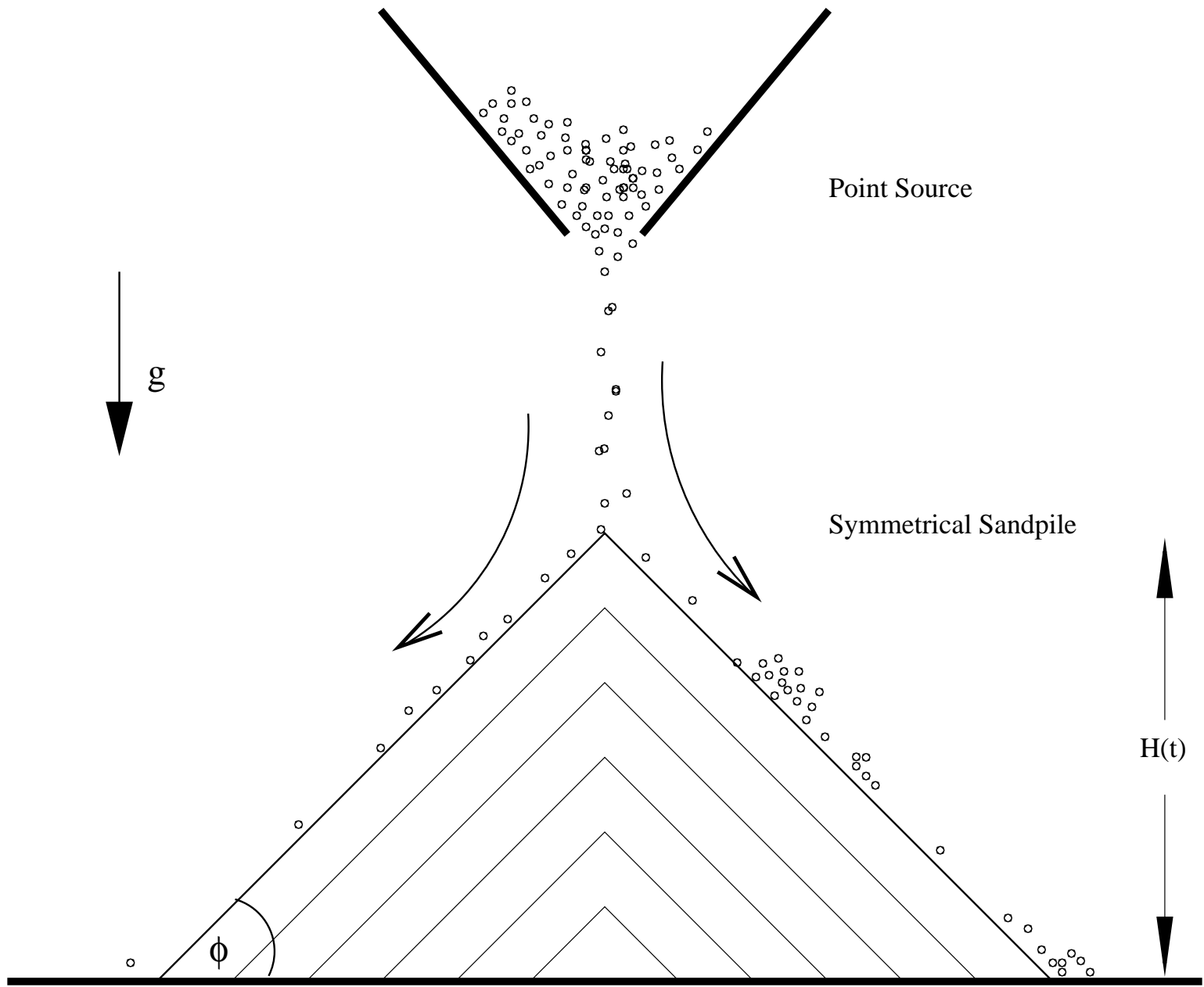
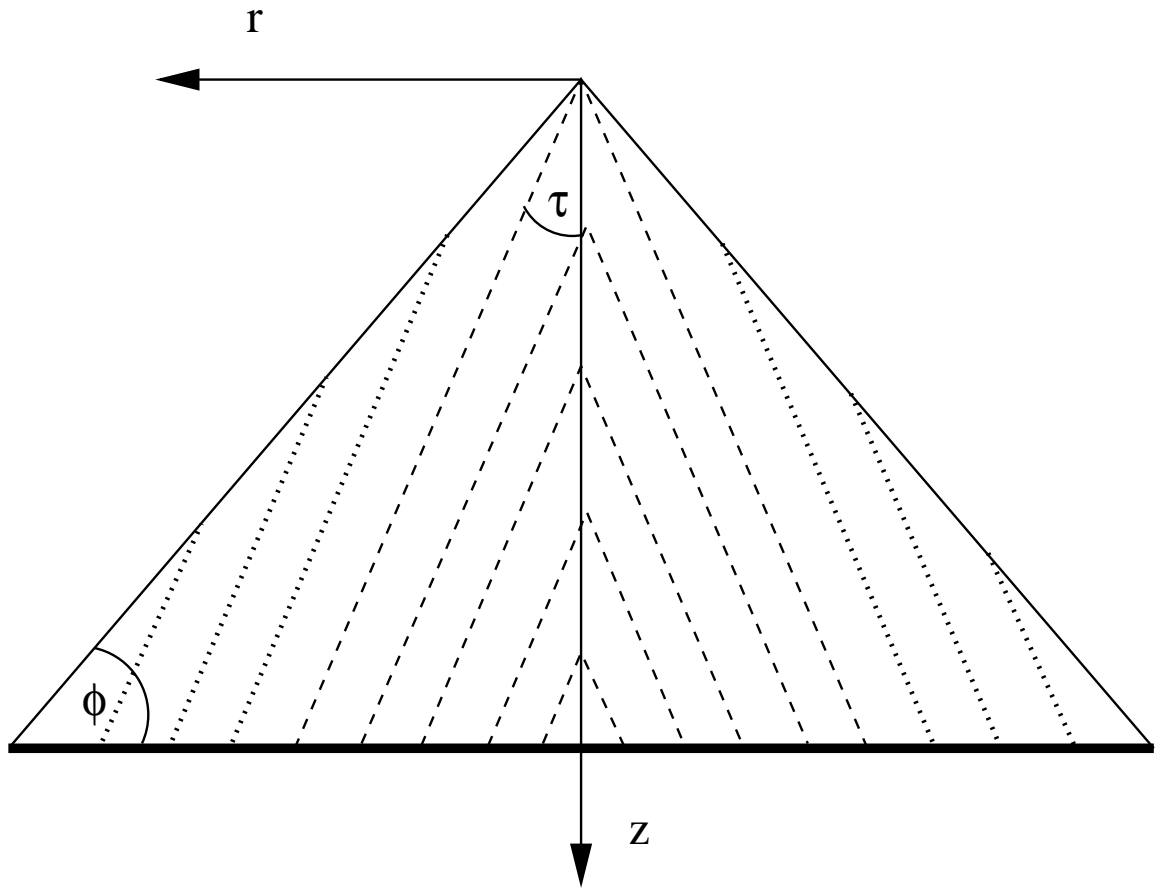
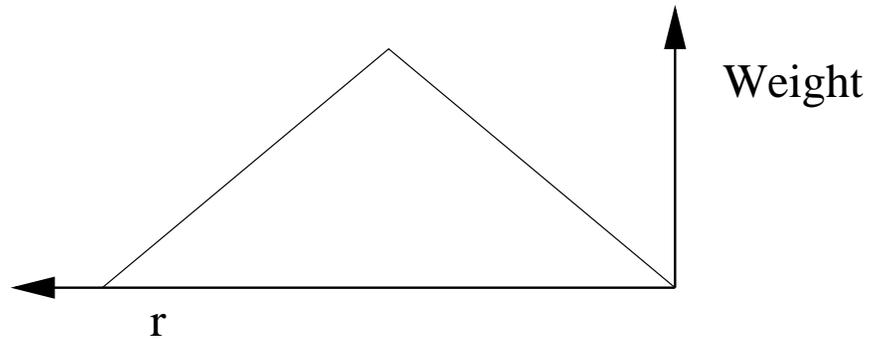
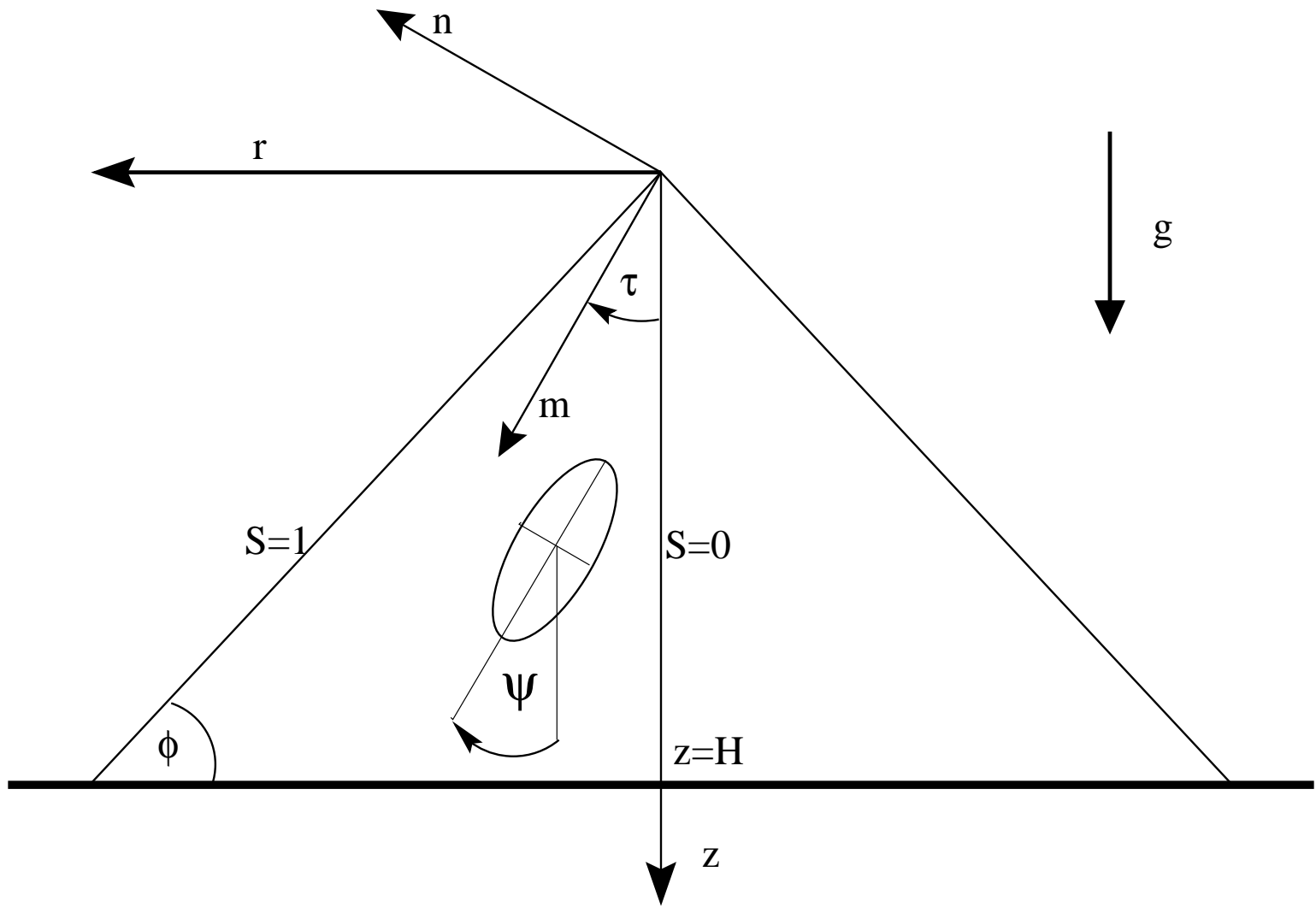
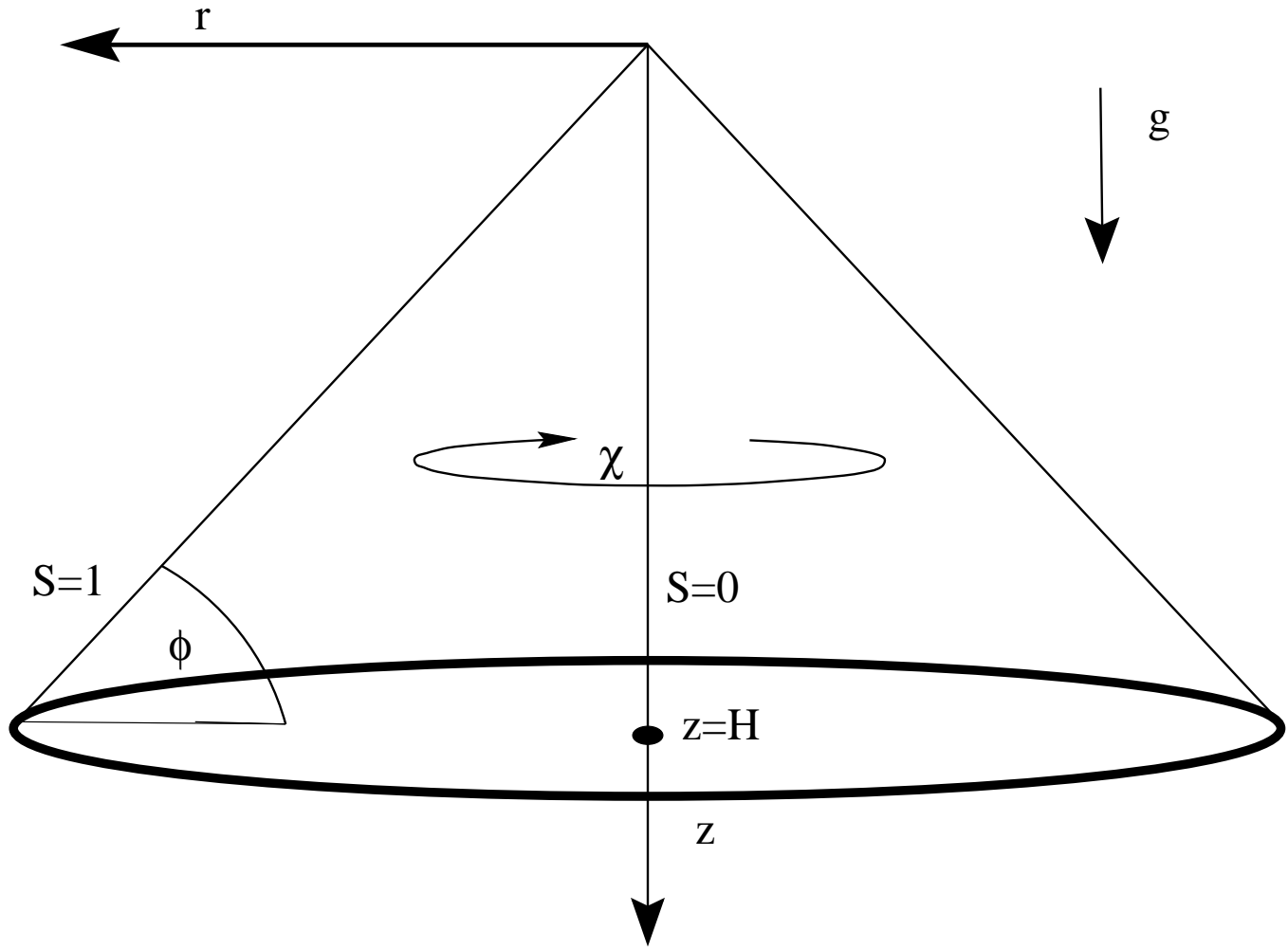
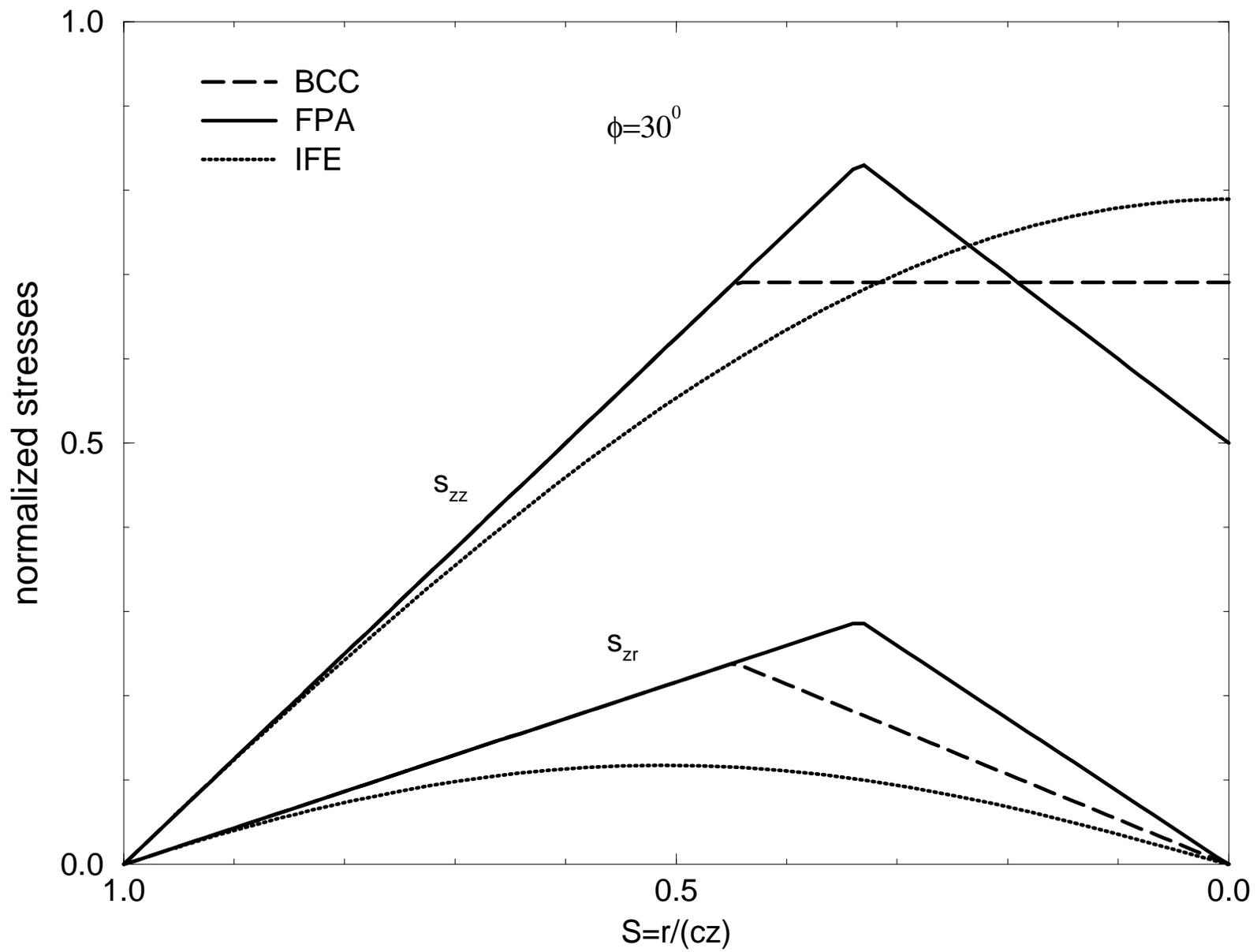
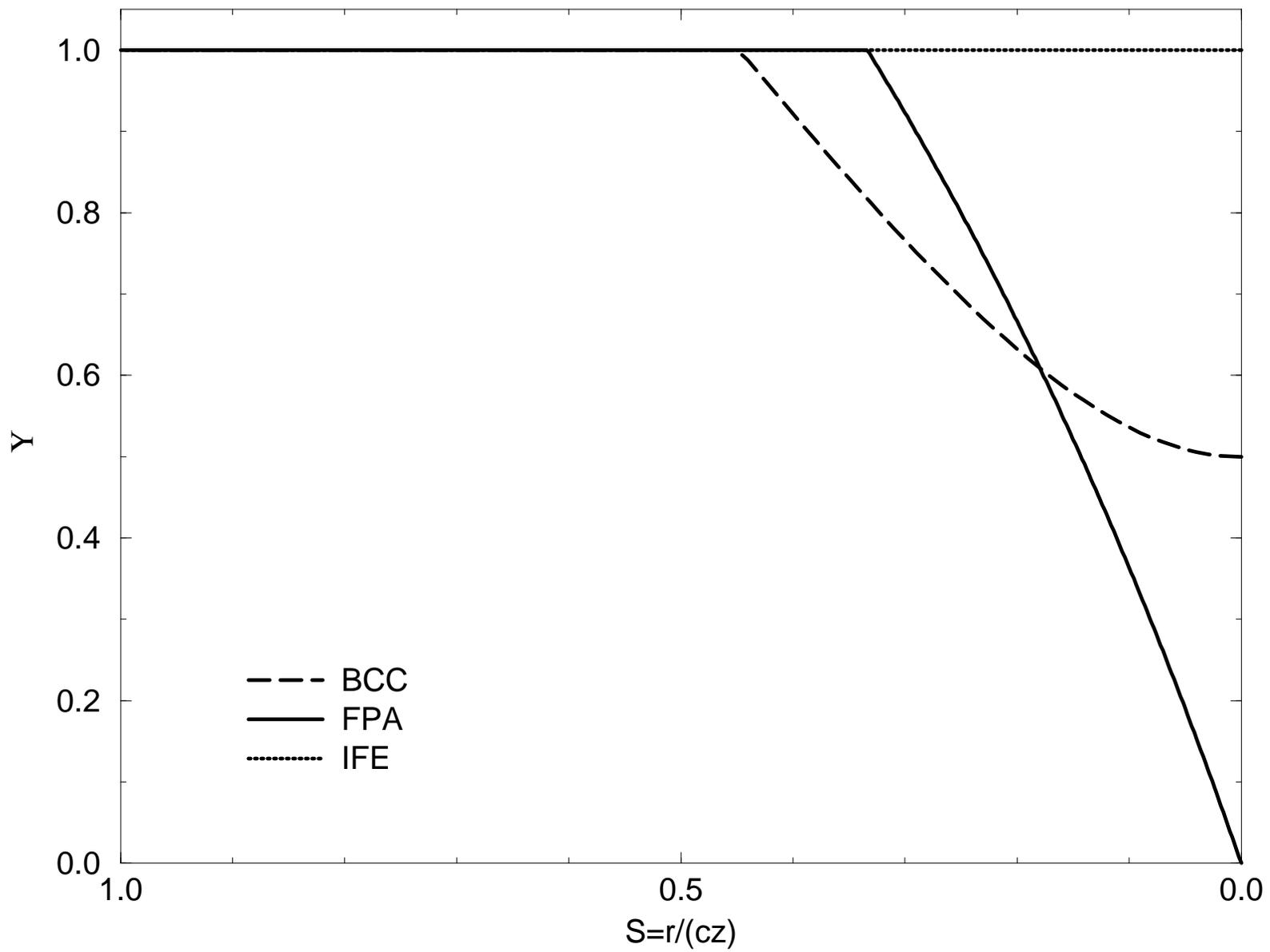
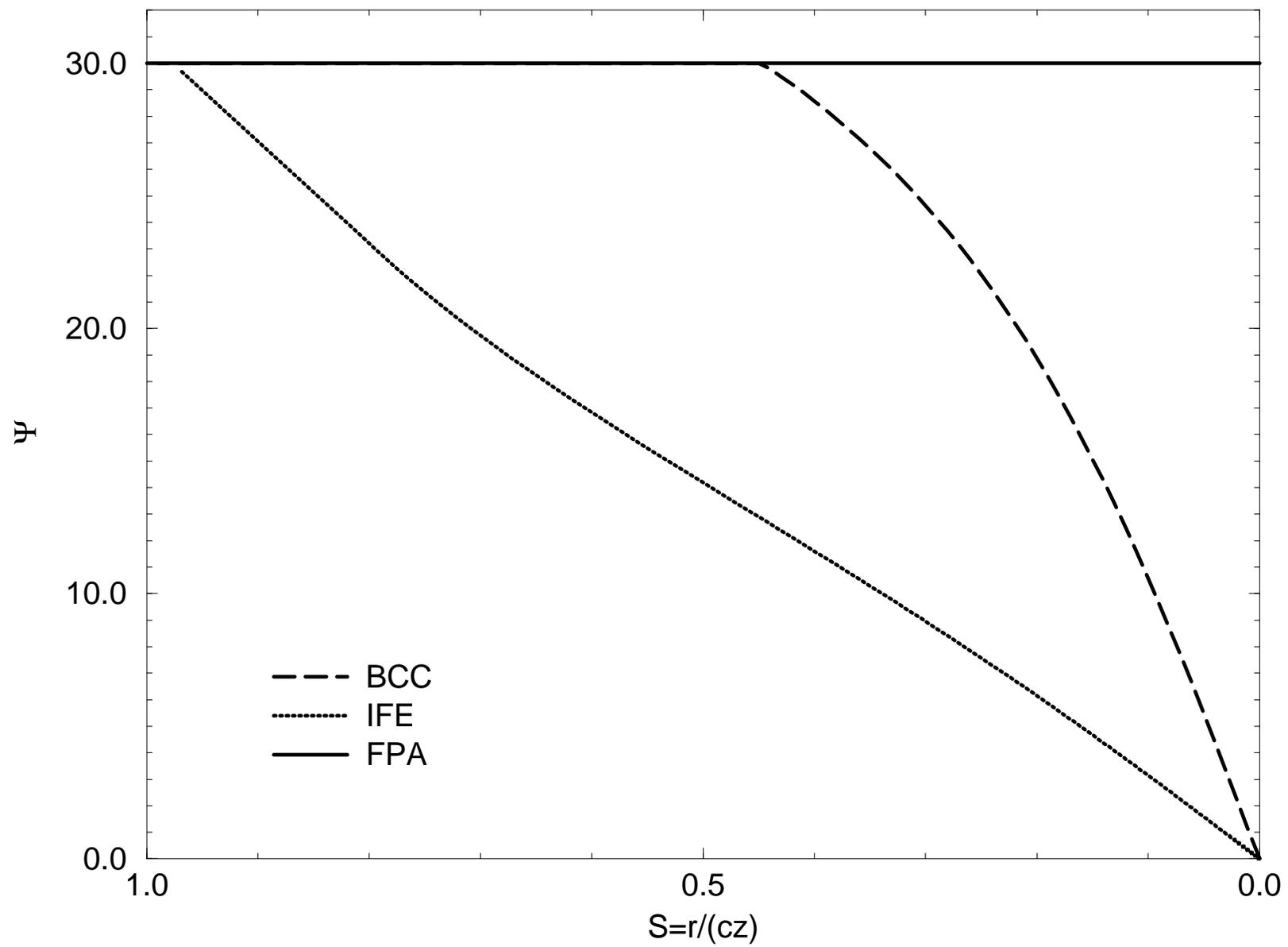
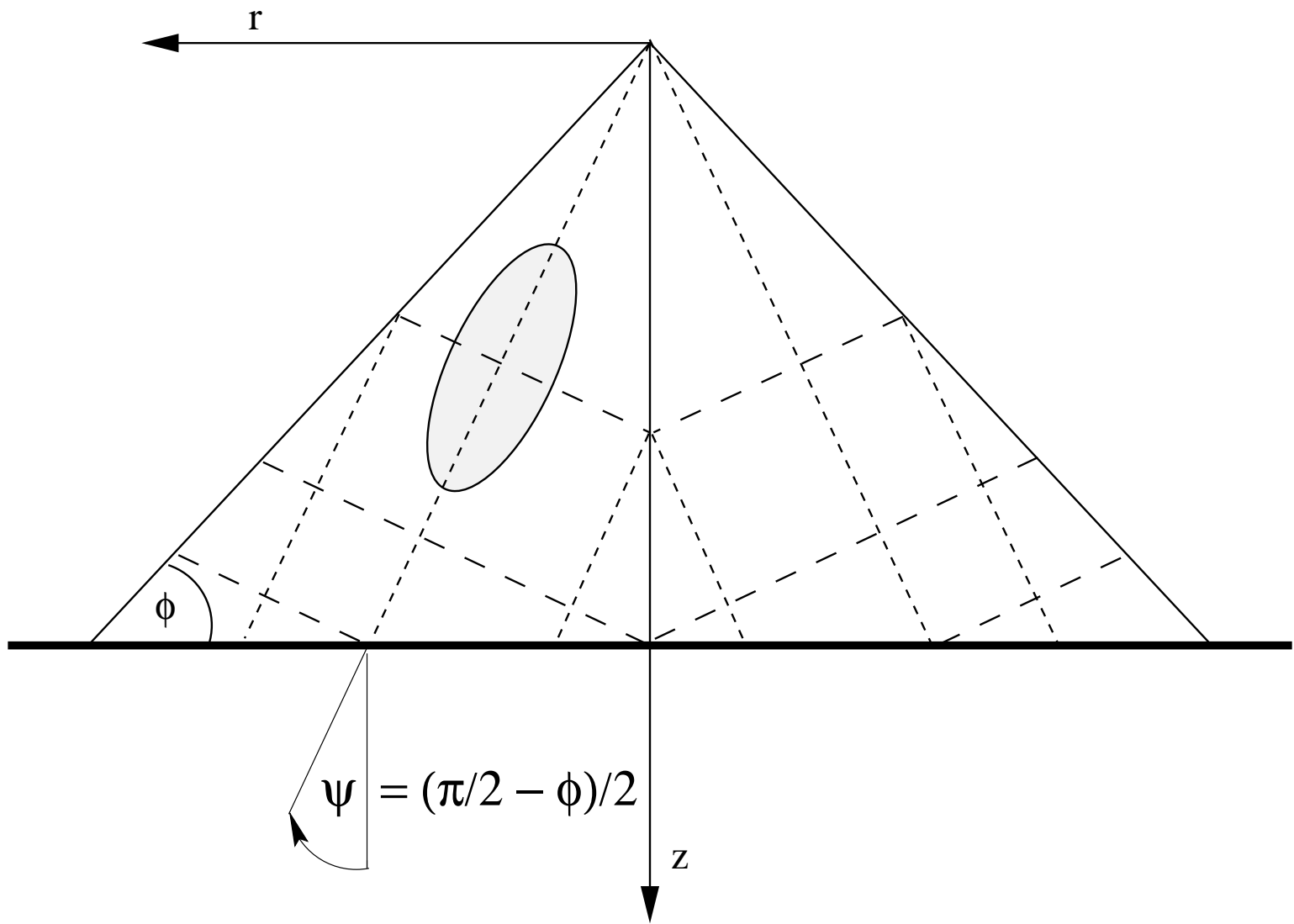
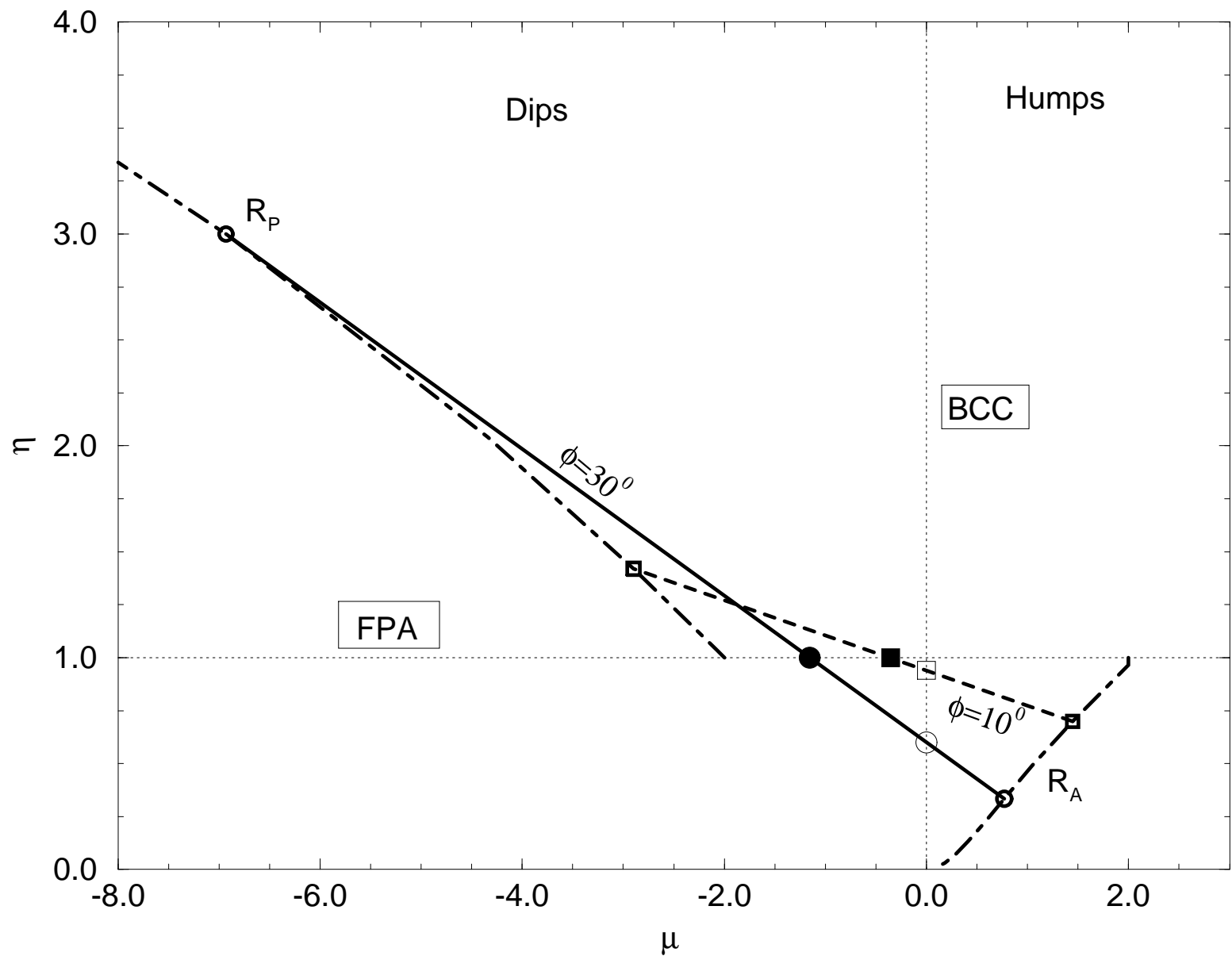
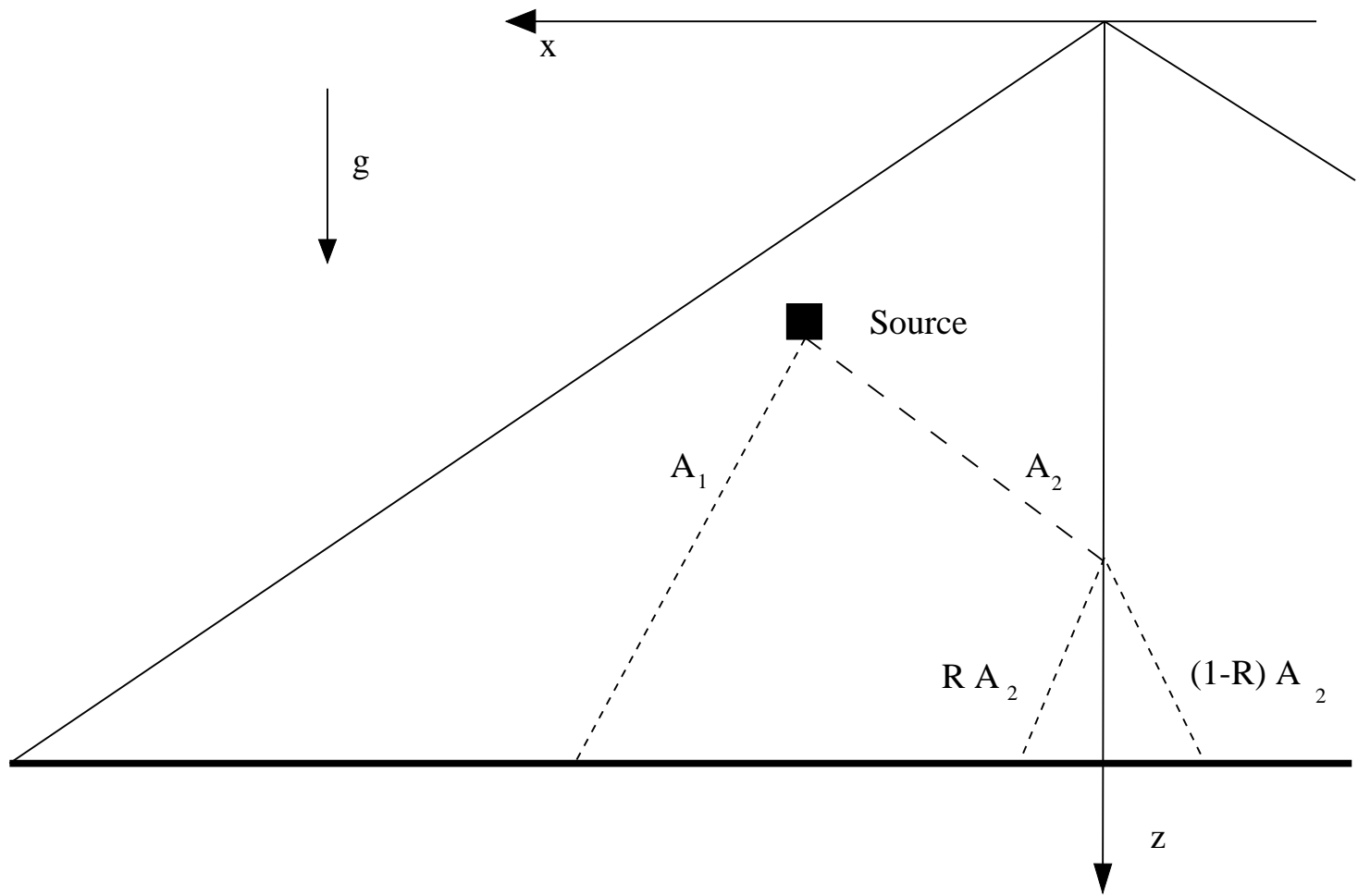
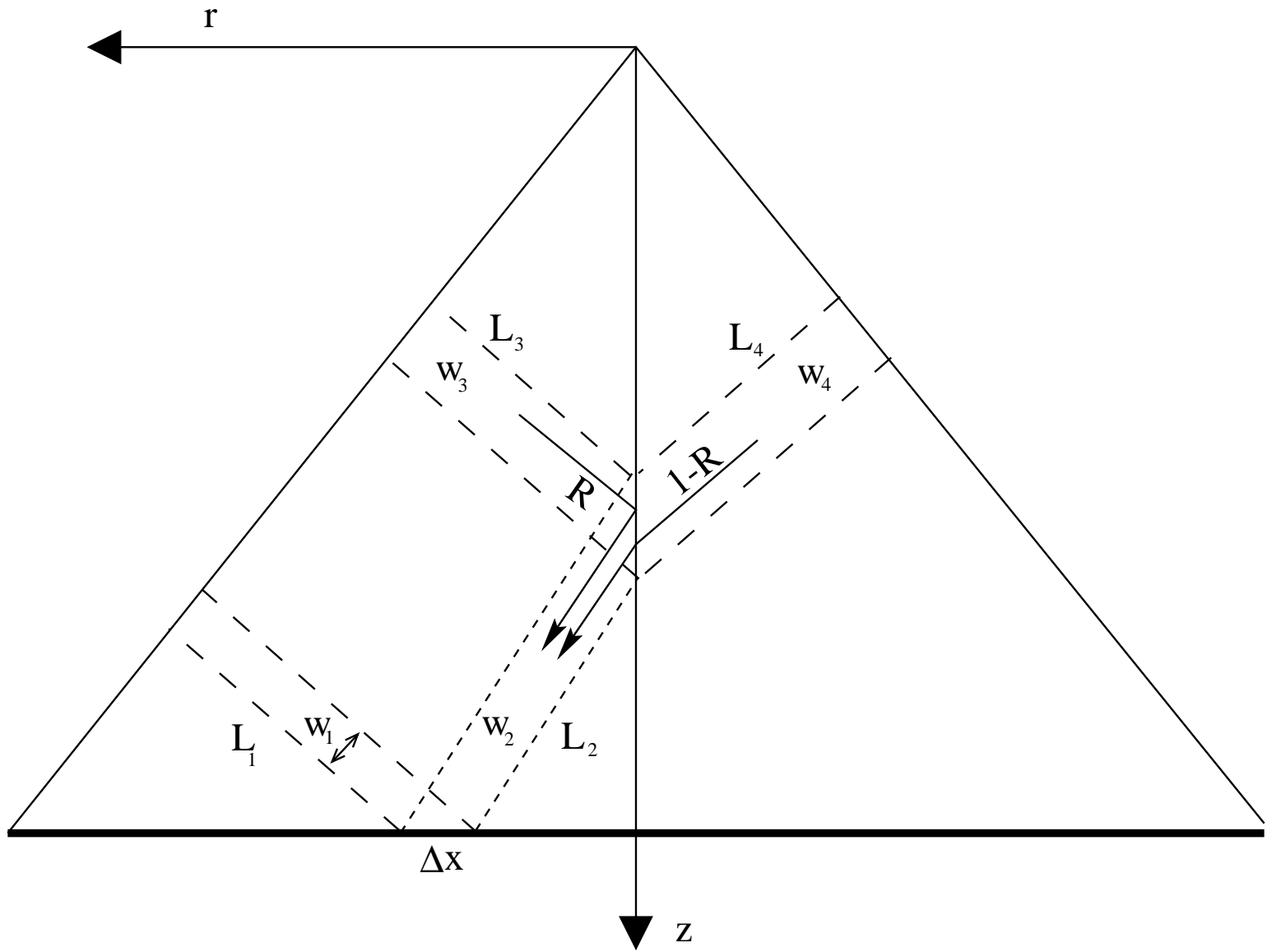
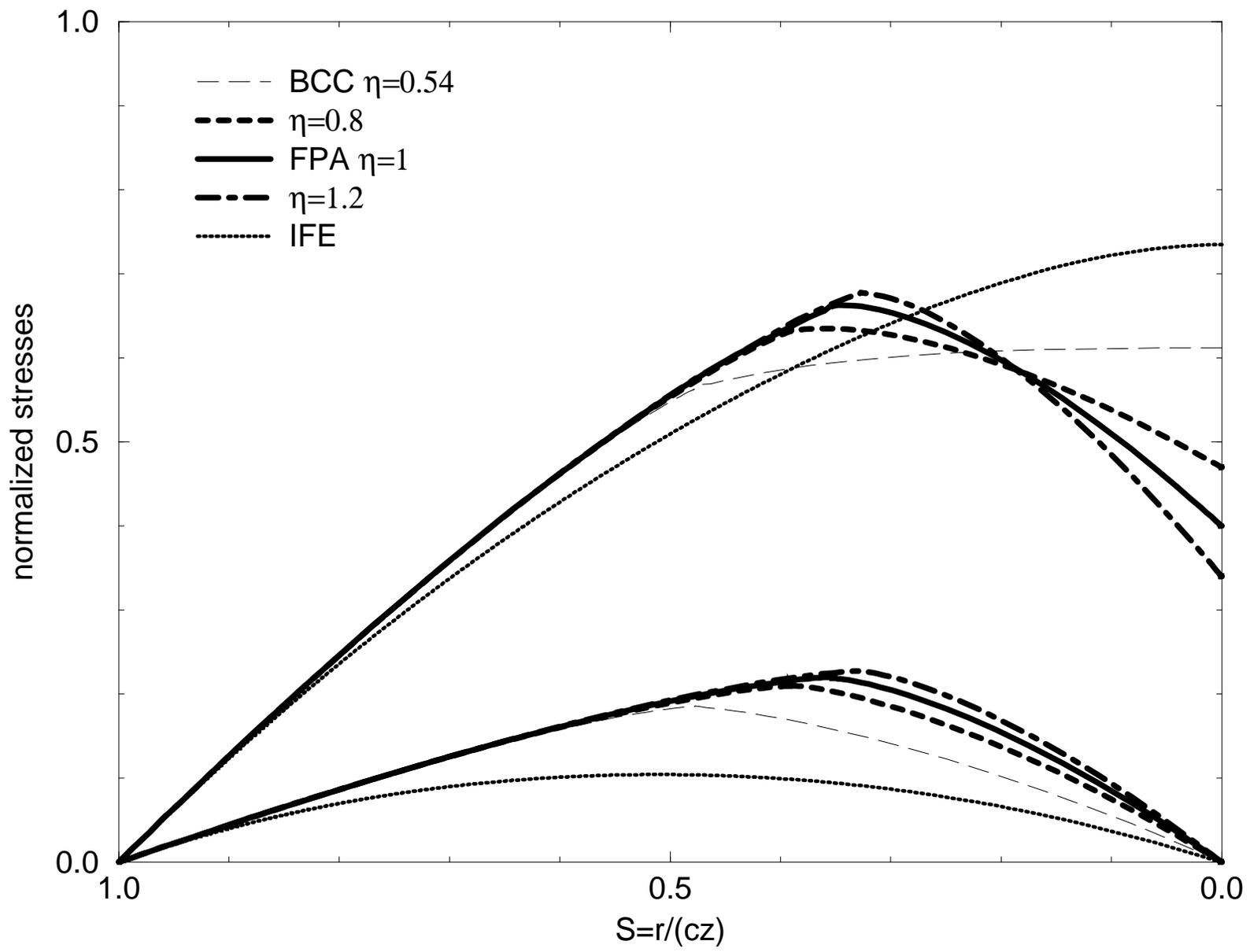
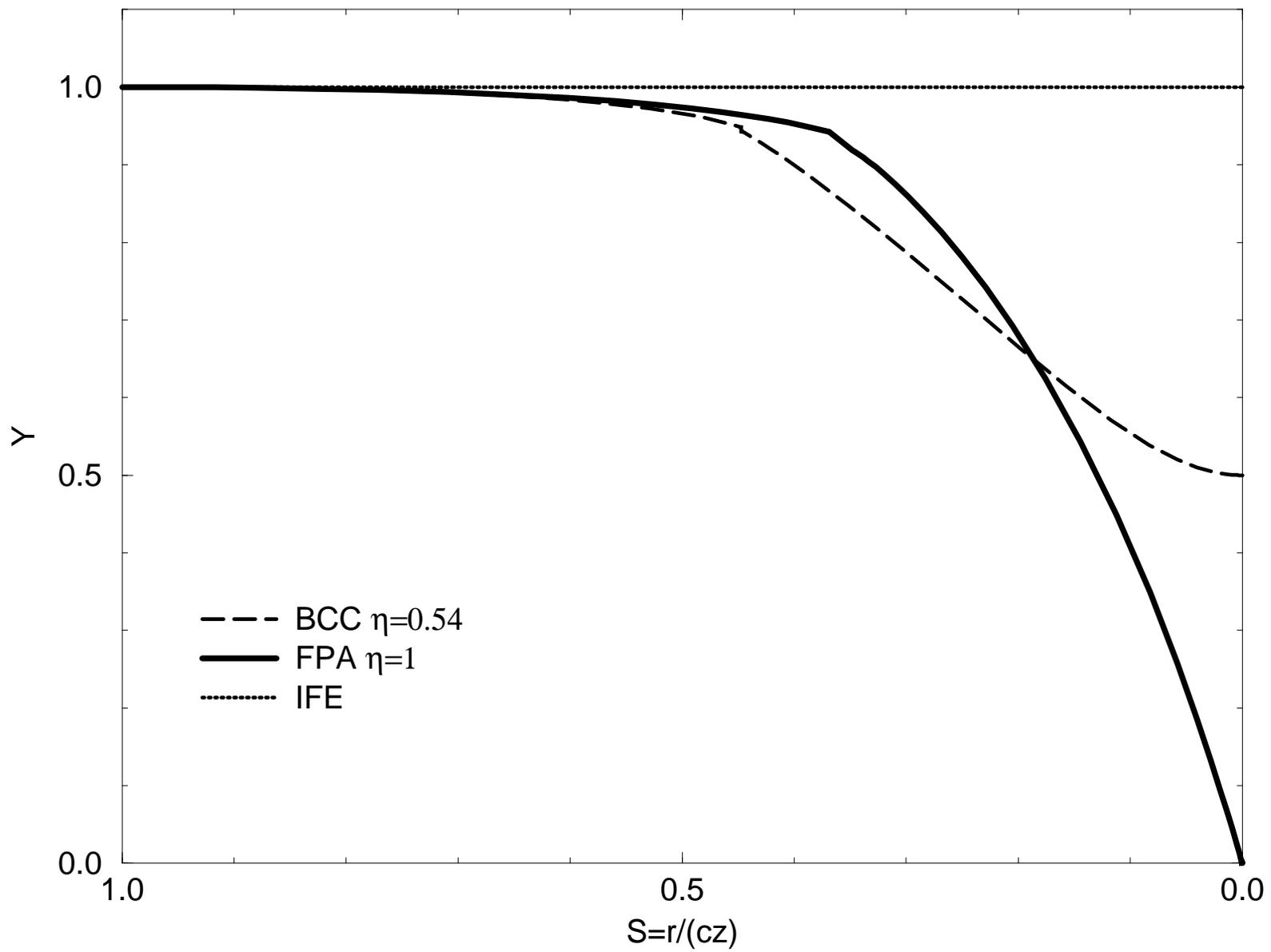
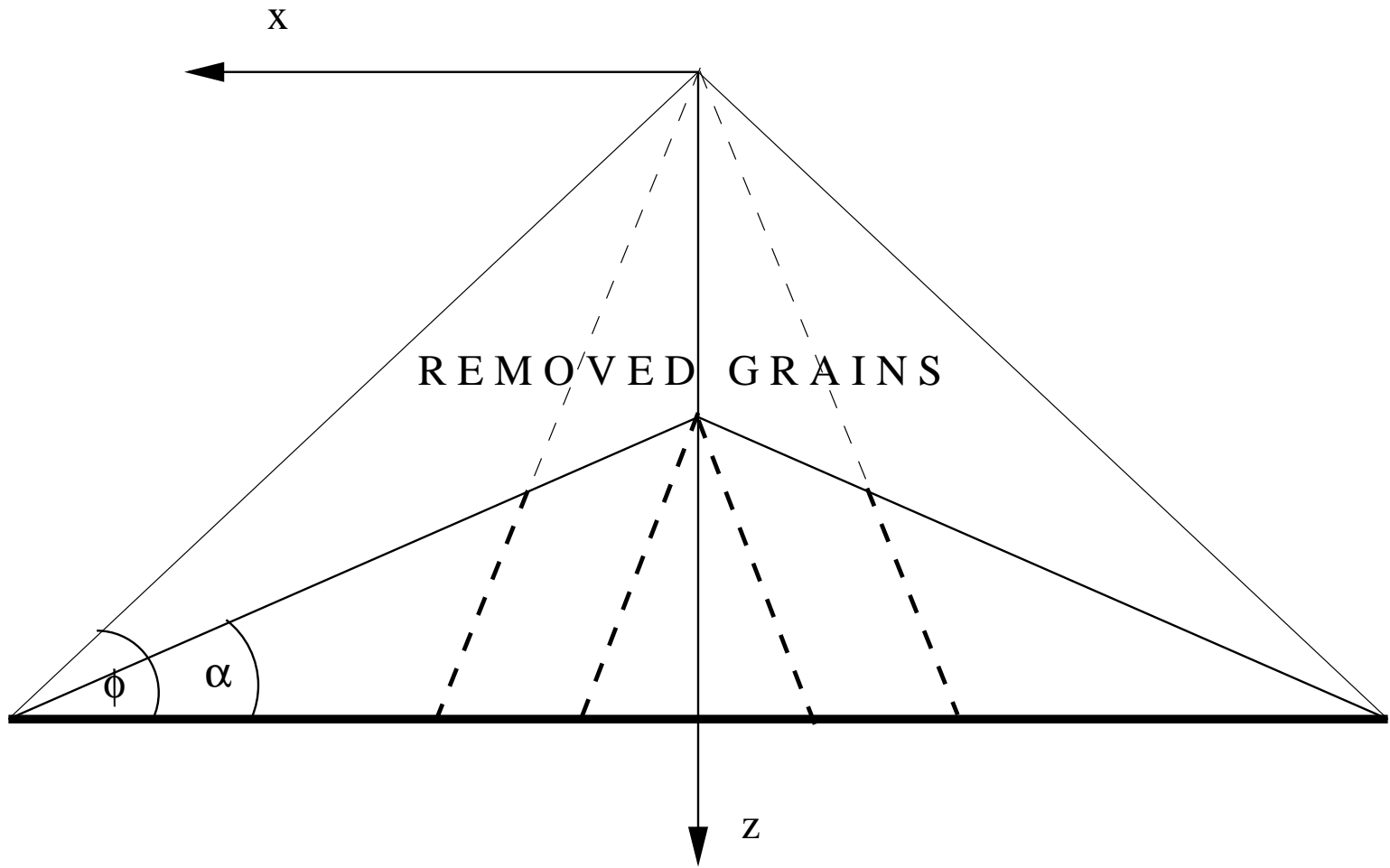
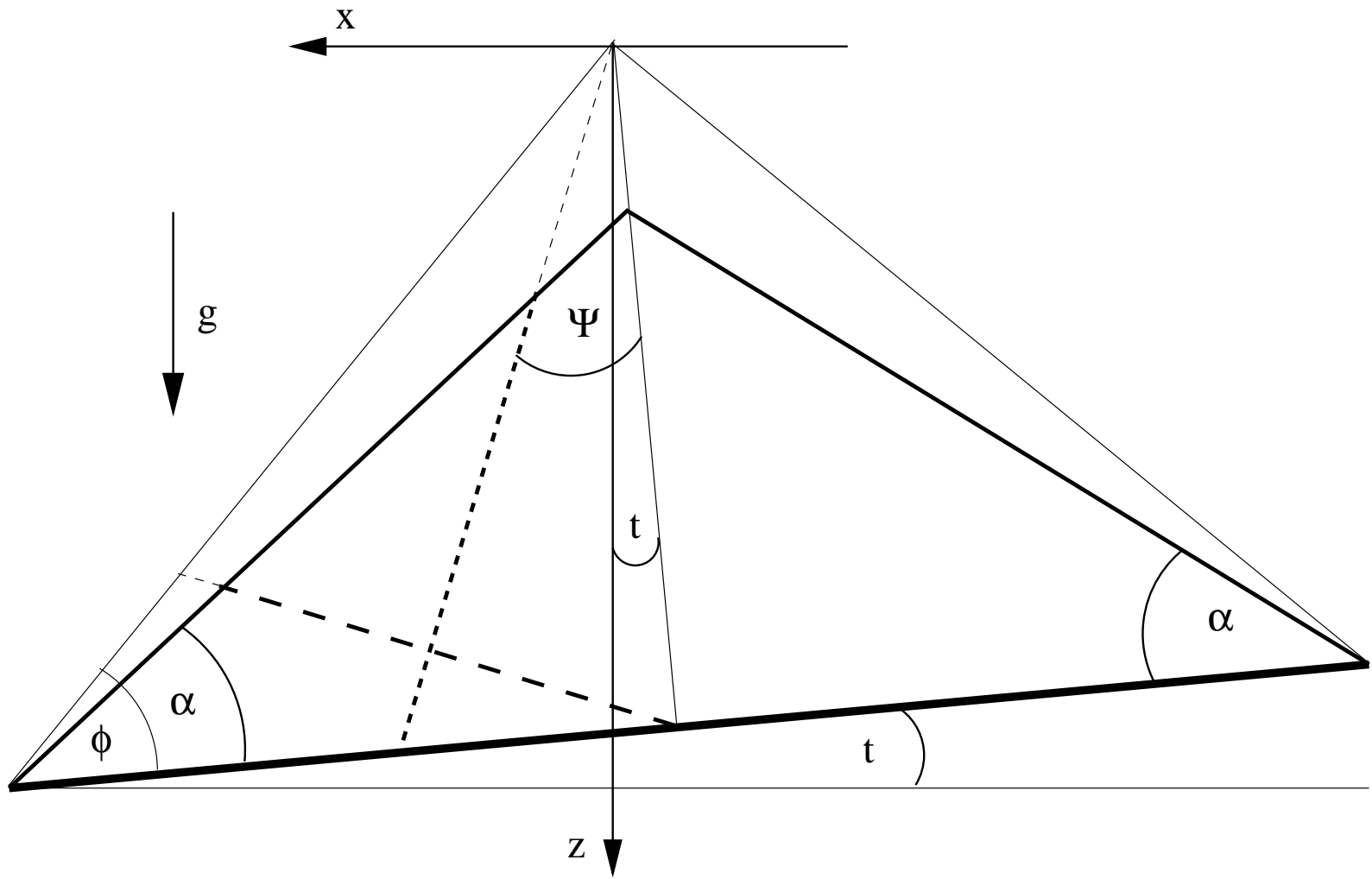
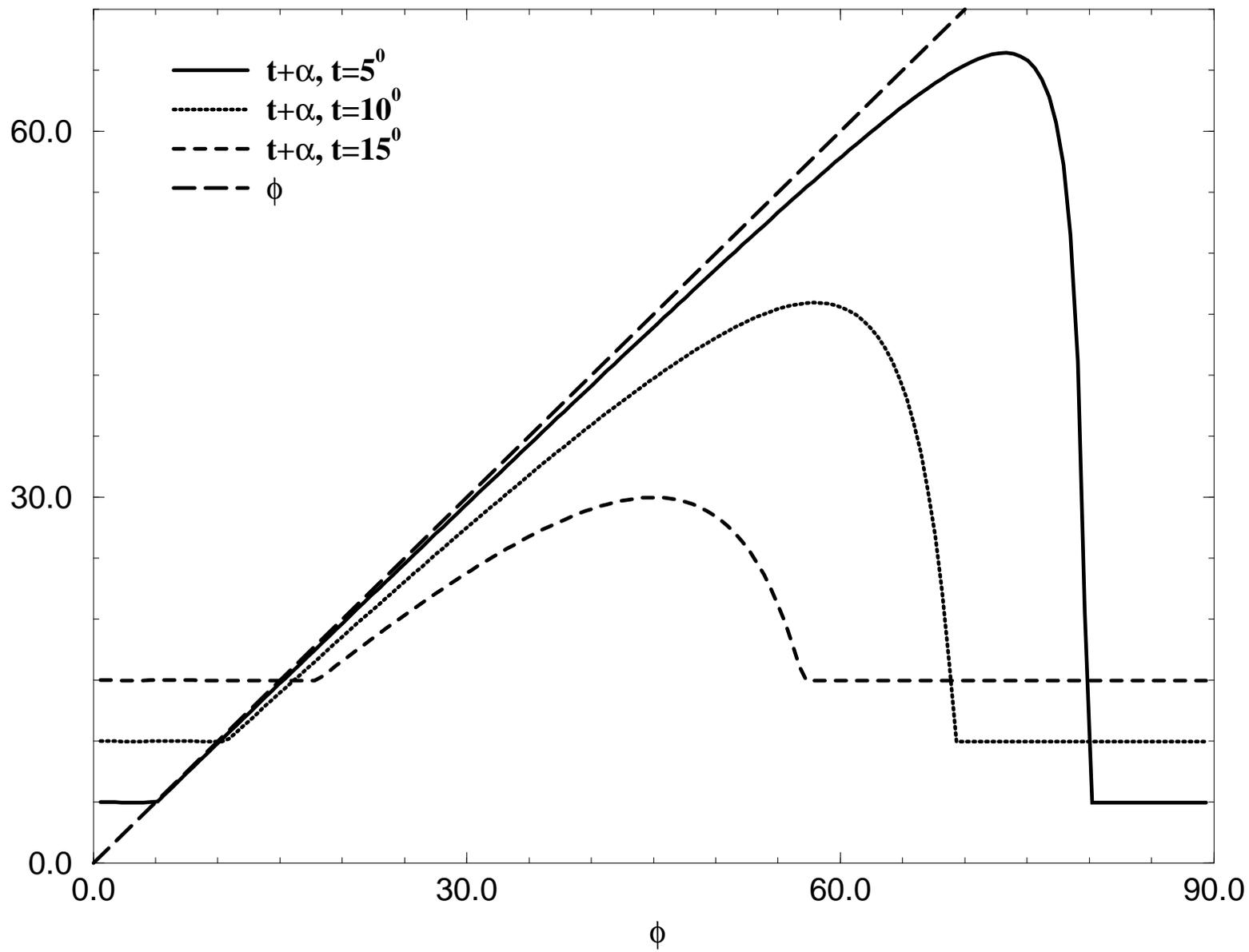
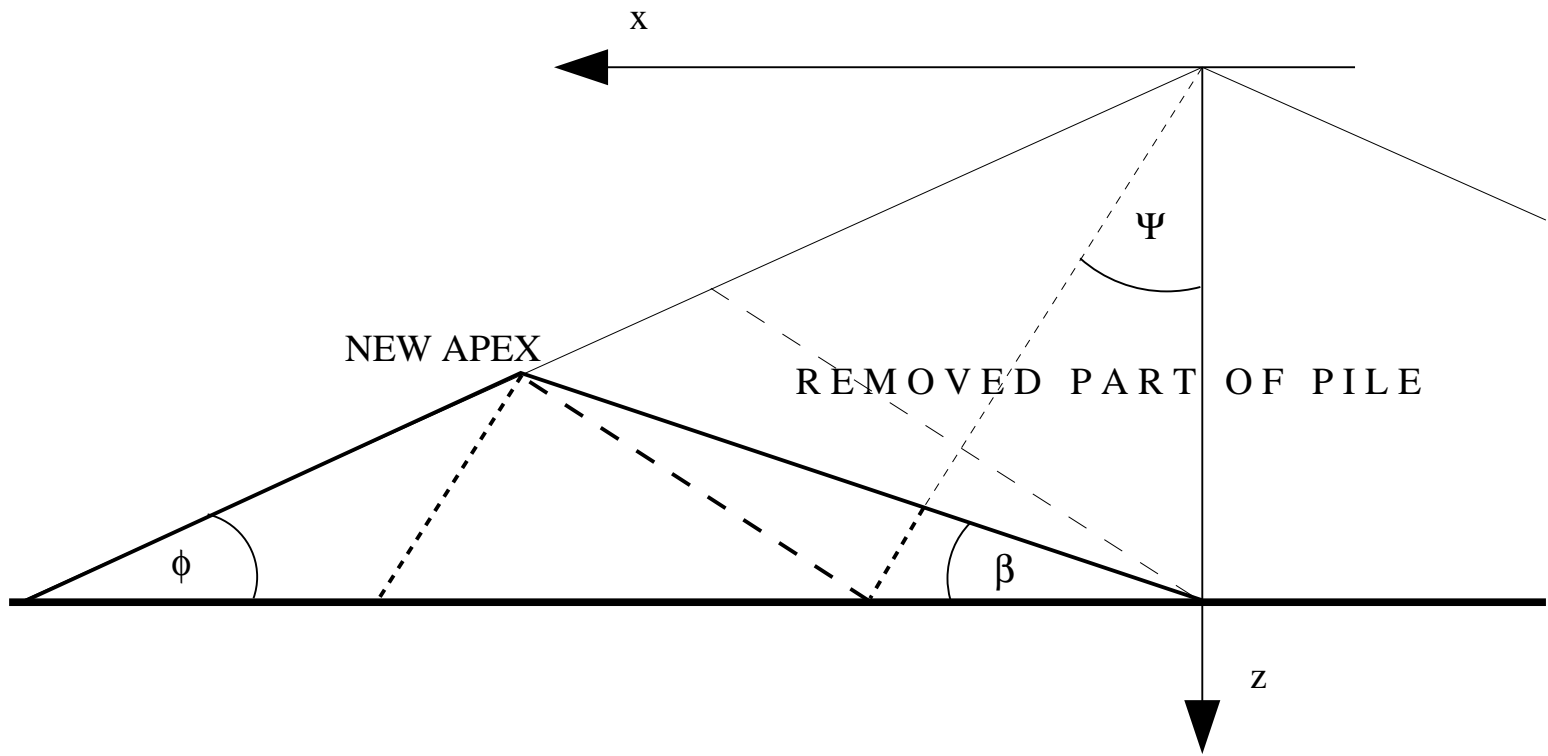
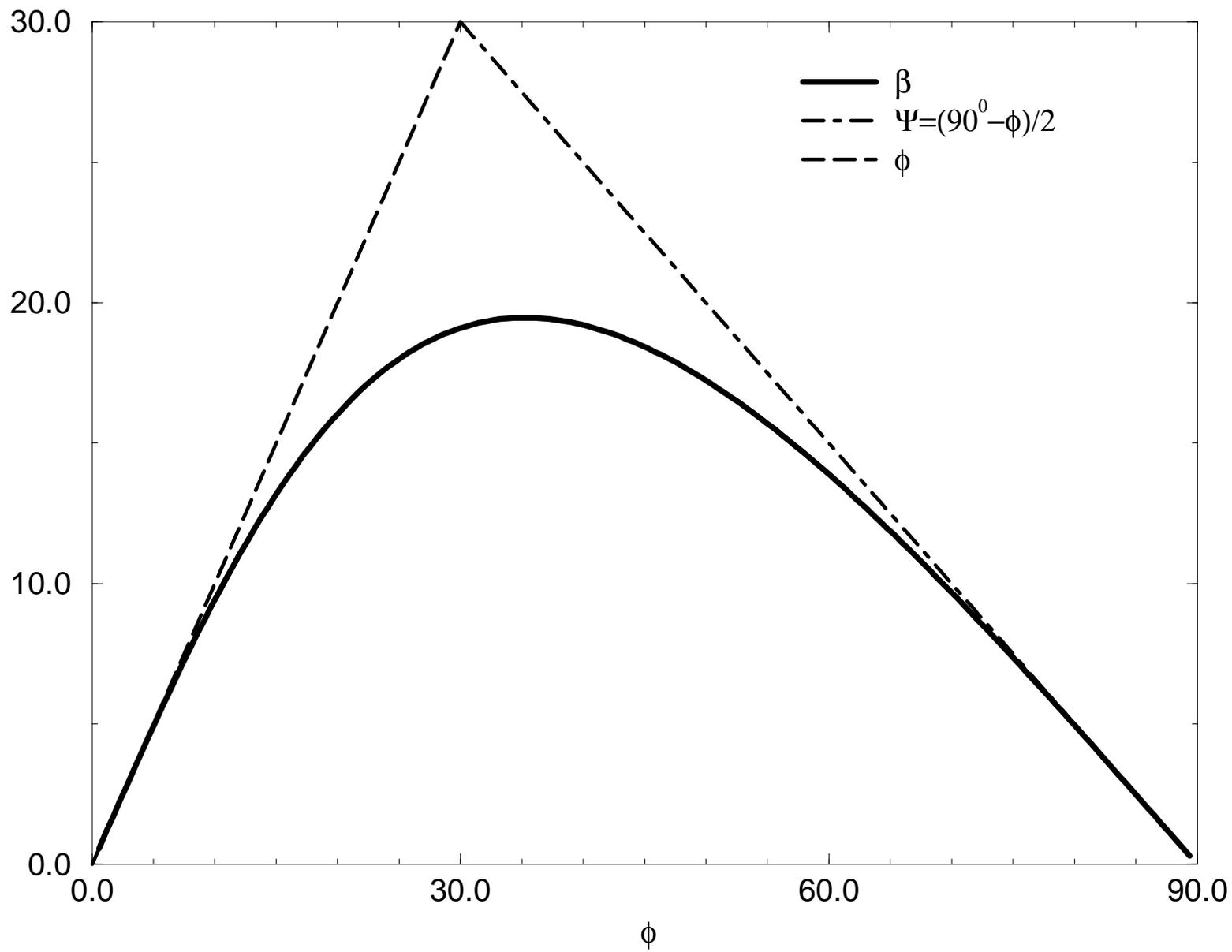
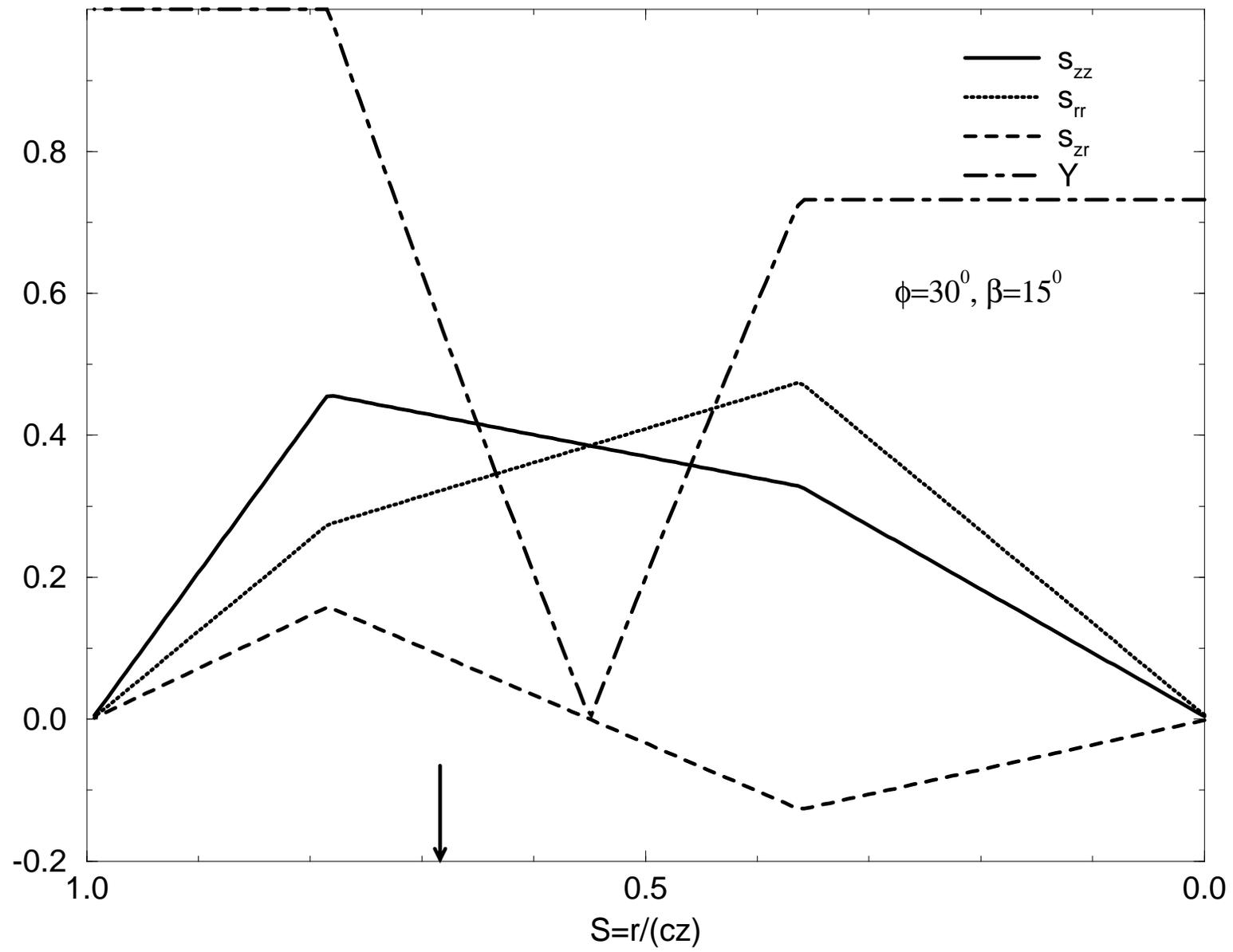
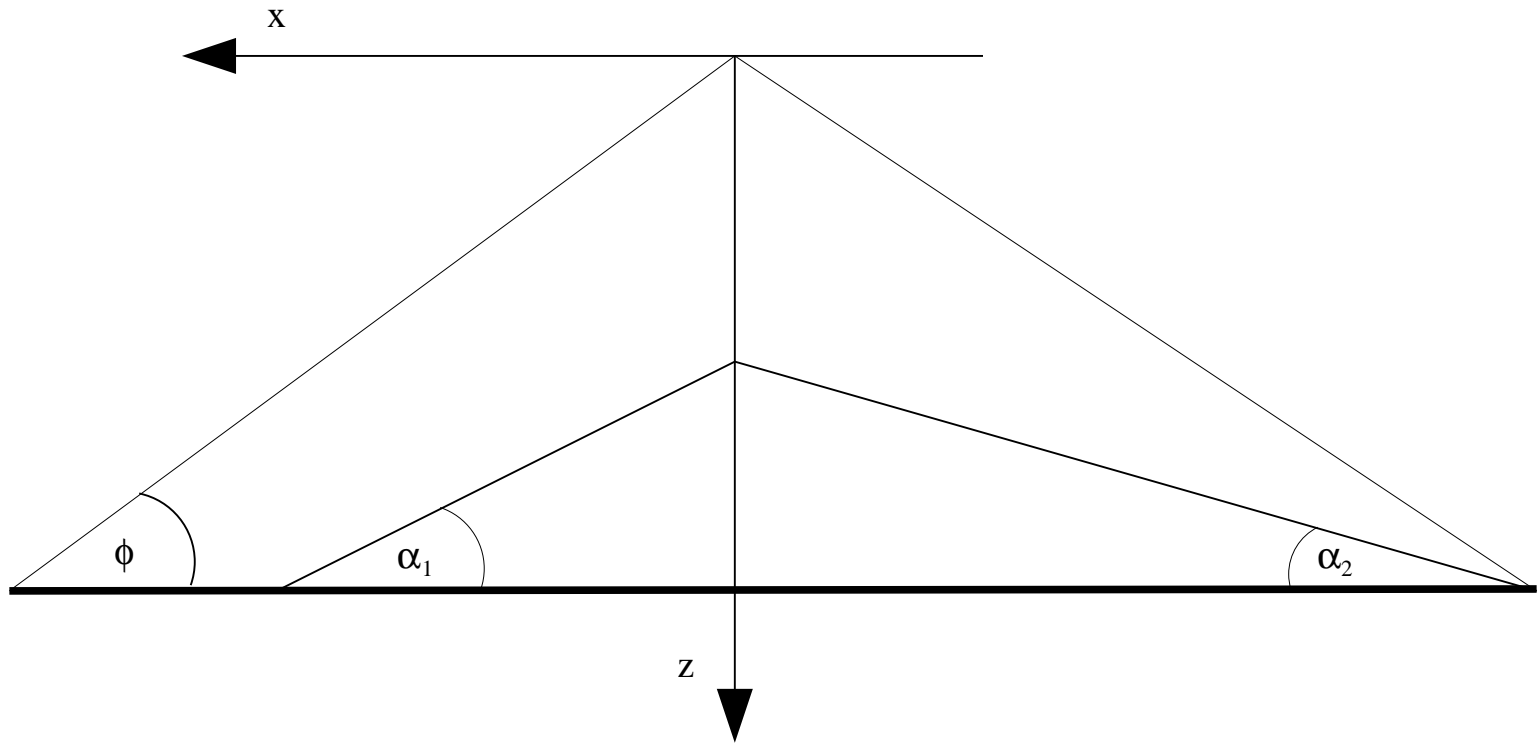
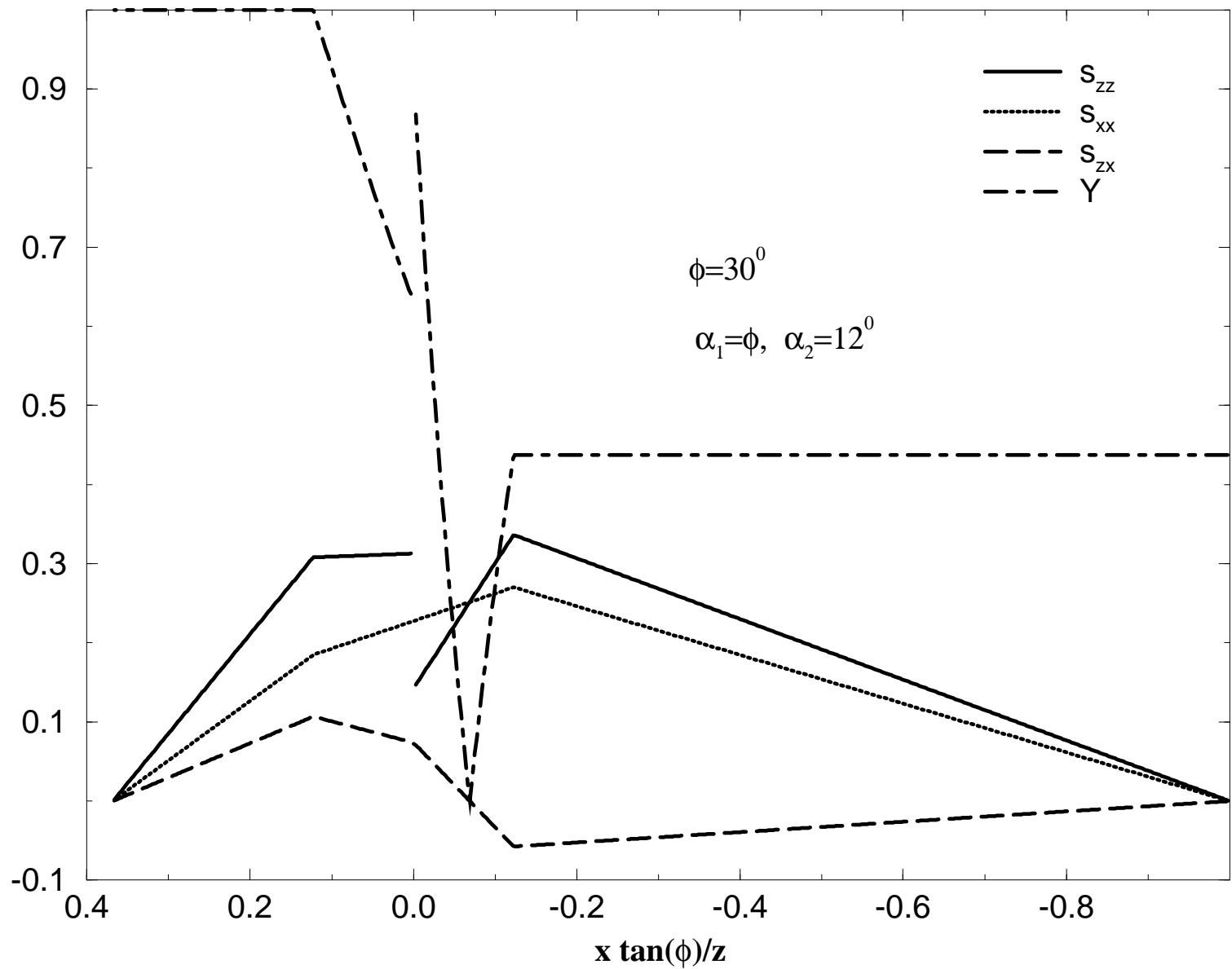
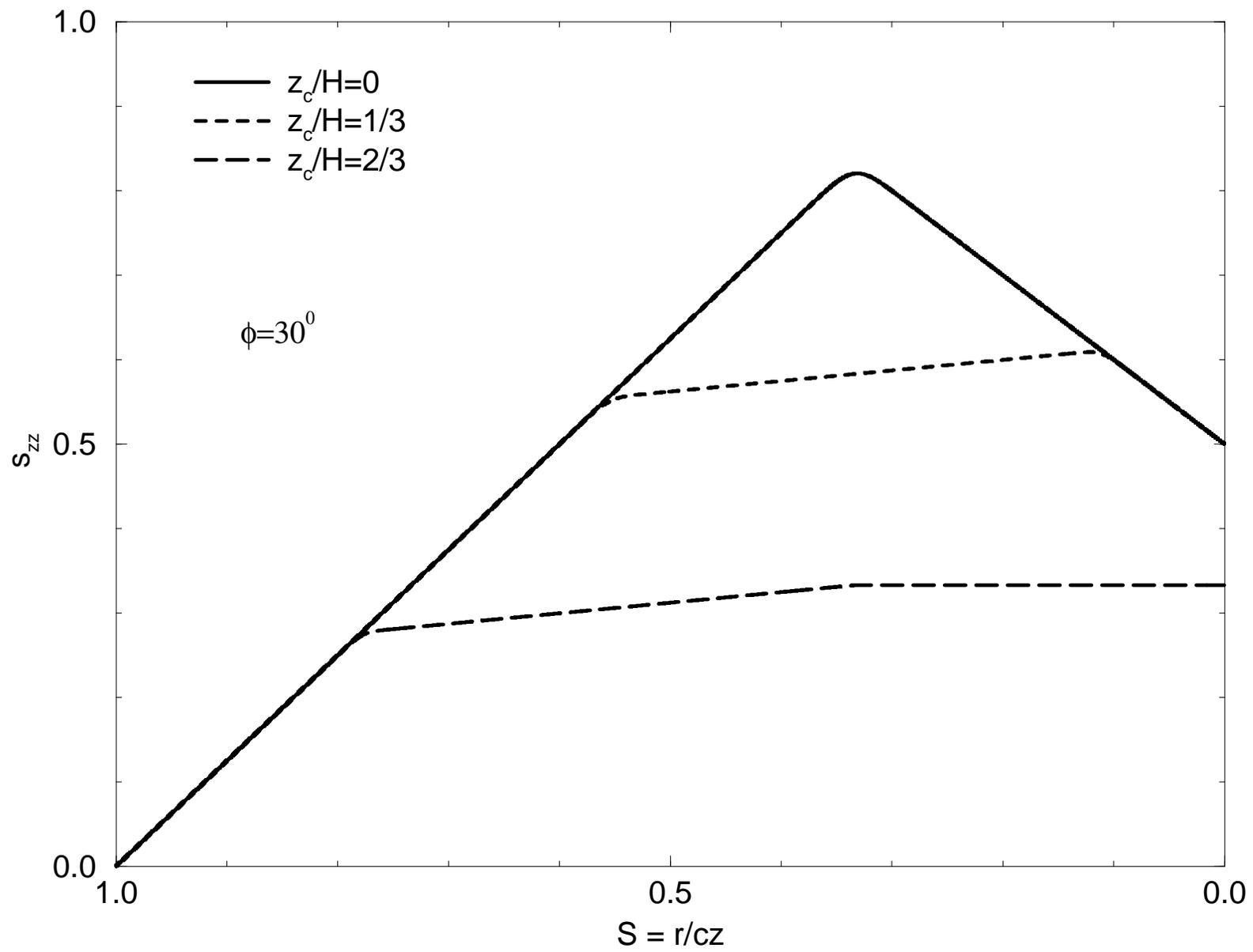
\end{document}